\documentclass[twocolumn,superscriptaddress,amsmath,aps,longbibliography,nofootinbib]{revtex4-1}
\usepackage{hyperref}
\usepackage{graphicx}
\usepackage{amsmath,amsfonts}
\usepackage{dcolumn}
\usepackage{bm}
\usepackage{color}
\usepackage{multirow}
\usepackage{float}
\usepackage{url}
\usepackage{physics}
\usepackage{subfigure}
\usepackage[utf8]{inputenc}
\usepackage{siunitx}

\begin{document}

\title{Superconductivity near the Mott-Ioffe-Regel limit in the high-entropy alloy superconductor (ScZrNb)$_{1-x}$(RhPd)$_x$ with a CsCl-type lattice}                      
\author{Sylwia Gutowska}
\affiliation{Faculty of Physics and Applied Computer Science,
AGH University of Science and Technology, Aleja Mickiewicza 30, 30-059 Krakow, Poland}

\author{Alicja Kawala}
\affiliation{Department of Quantum Many-Body Theory, Jagiellonian University, ul. prof. S. Łojasiewicza 11, 30-348 Krakow, Poland}

\author{Bartlomiej Wiendlocha}
\email{wiendlocha@fis.agh.edu.pl}
\affiliation{Faculty of Physics and Applied Computer Science,
AGH University of Science and Technology, Aleja Mickiewicza 30, 30-059 Krakow, Poland}

\date{\today}

\begin{abstract}
Theoretical analysis of the electronic structure of the high-entropy-type superconductor (ScZrNb)$_{1-x}$(RhPd)$_x$, $x \in (0.35, 0.45)$ is presented. 
The studied material is a partially ordered CsCl-type structure, with two sublattices, randomly occupied by Sc, Zr, Nb (first sublattice) and Nb, Rh, and Pd (second sublattice). Calculations were done using the Korringa-Kohn-Rostoker method with the coherent potential approximation (KKR-CPA) and take into account the substitutional disorder. Our total energy calculations confirm the preference for the partially ordered structure over the fully random {\it bcc}-type one. Electronic densities of states $N(E)$, dispersion relations, and McMillan-Hopfield parameters $\eta$ (electronic contribution to electron-phonon coupling) are studied as a function of composition. 
The computed increasing trends in $N(E_F)$ and $\eta$ with $x$ are opposite to what we expected based on the experimental results, where the decrease in the critical temperature with increasing $x$ was found. Very strong electron scattering due to disorder is observed, as the electronic dispersion relations are strongly smeared. As a result, the computed electronic lifetimes $\tau$ are very short, leading to a small mean-free path of electrons of the order of interatomic distance, which puts (ScZrNb)$_{1-x}$(RhPd)$_x$ near the Mott-Ioffe-Regel limit. The trend in $\tau(x)$ is similar to the trend observed experimentally in $T_c(x)$, suggesting that disorder may be the factor that influences $T_c$ in this series of alloys. 
\end{abstract}


\maketitle

\section{Introduction}
High entropy alloys (HEAs) are usually defined as alloys containing five or more elements that randomly occupy lattice positions in simple "monoatomic" crystal structures~\cite{yeh_2004,yeh_2007}.
The concentration of elements varies between 5\% and 35\%, and the most common structures are the simplest cubic body centered {\it bcc} (tungsten-type) or face centered {\it fcc} (cooper-type) ones.
The configuration entropy plays an important role in their formation; hence, the name. HEAs exhibit interesting properties from both a functional and a fundamental scientific point of view~\cite{yeh_2004,yeh_2007,ZHANG2014,tsai2014,MIRACLE2017,george2019,murty2019}.

Several superconducting HEAs have been found so far~\cite{kozejl_2014,tnhzt1,tnhzt2,hea-sup1,prb_pikul}, with the first one, Ta$_{0.34}$Nb$_{0.33}$Hf$_{0.08}$Zr$_{0.14}$Ti$_{0.11}$~\cite{kozejl_2014} 
(TNHZT in short) reported in 2014. 
TNHZT crystallizes in the above-mentioned $Im$-$3m$ {\it bcc}-type of structure, with all atoms randomly occupying the crystal site (2a). It has a superconducting transition temperature of $T_c = 7.3$~K. 
Due to the flexibility in tuning the average number of valence electrons per atom, one can investigate its relation to the critical temperature, testing the classical Matthias rule~\cite{matthias} (maximum $T_c$ for about 5 and 7 valence electrons per atom). Although simple {\it bcc} HEAs were found to follow the rule~\cite{tnhzt1,tnhzt2}, different behaviors were observed in more complex systems~\cite{stolze2018,hea-cscl-cava}.

Due to their structural complexity, HEAs should offer a unique opportunity to investigate the interplay of disorder and superconductivity.
Since Anderson's theorem ~\cite{anderson_59} we know that superconductivity in a conventional superconductor is generally robust with respect to the presence of non-magnetic defects and weak disorder. However, this does not apply to unconventional superconductors, with a good example of Sr$_2$RuO$_4$~\cite{sr2ruo4-disorder}, where $T_c$ is quickly suppressed by non-magnetic impurities. 
This occurs when the electronic mean-free path is large, $d > 500$\AA, that is, hundreds of interatomic distances. 
In contrast, the disorder was found to increase $T_c$ in some untypical cases, e.g., in a cuprate La$_{1.875}$Ba$_{0.125}$CuO$_4$ ~\cite{disorder_raises_cuprate} or
in monolayer Nb$_2$Se~\cite{raises_nbse2}, but this is a rare situation. 

In highly disordered materials, on the other hand, disorder can suppress superconductivity~\cite{anderson_degradation,weakly_localized_regime}. This was observed in A-15 superconductors~\cite{anderson_degradation}, in granular and highly disordered metals~\cite{destruction_granular}, or in thin films~\cite{destruction_films}.
Not only $T_c$ may be affected by disorder. For two-band anisotropic superconductors, 
in addition to the reduction of $T_c$ with a decrease in electronic lifetime $\tau$, 
the specific heat jump across the superconducting transition $\Delta C/\gamma T_c$ ($\gamma$ is the Sommerfeld coefficient) was found to be reduced below the BCS value of 1.43~\cite{heat-prb}.
Renormalization of electron-phonon coupling was reported in the V$_{1-x}$Ti$_x$ alloy superconductors~\cite{renormalization_elph}, whe\-re due to the presence of point defects, the Mott-Ioffe-Regel limit (mean-free path of electrons becoming as short as the interatomic distance) was reached.

To put the results of our current work into the proper context, we will briefly review the most important results obtained so far for Ta$_{0.34}$Nb$_{0.33}$Hf$_{0.08}$Zr$_{0.14}$Ti$_{0.11}$., the first superconducting HEA.
Although it is a highly disordered material with five elements occupying a single site in the unit cell, 
the electronic structure calculations~\cite{jasiewicz_2016} gave the quite unexpected result that disorder has a minor impact on its electronic band structure.
This issue was studied using the complex band technique within the Korringa-Kohn-Rostoker method with the coherent potential approximation (KKR-CPA)~\cite{butler_1985}. 
Disorder is a source of electron scattering, and the magnitude of this effect can be quantified by calculating the
imaginary part of the energy, ${\rm Im}(E)$, as energy is a complex variable in KKR-CPA. 
The imaginary part describes the disorder-induced band sme\-a\-ring effect, 
the stronger is the electron scattering, the more smeared are the bands (the larger the ${\rm Im}(E)$), and the smaller the electronic lifetime $\tau$~\cite{butler_1985}:
\begin{equation}\label{eq:tau}
    \tau = \frac{\hbar}{2{\rm Im}(E)}.
\end{equation}
In TNHZT, despite the complete substitutional disorder, calculations \cite{jasiewicz_2016} showed that there are sharp, well-defined electronic bands, with a minor disorder-induced band smearing effect (see also the discussion in Sect. \ref{sec:bands}). Electron scattering appeared to be relatively weak and not stronger than in typical binary alloys. Thus, from a band structure point of view, the TNHZT superconductors do not behave as strongly disordered materials.

\begin{figure}[t]
 \centering
\includegraphics[width=0.95\columnwidth]{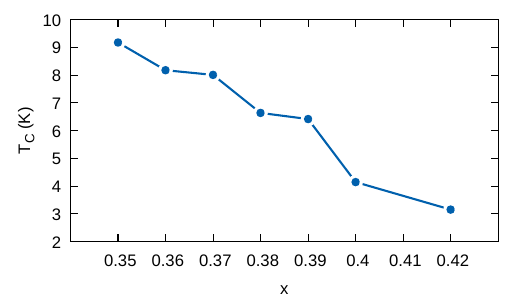}
\caption{\label{tc_x}Experimentally determined critical temperature $T_c$ (points, lines are added to highlight the trend) as a function of the alloy composition, $x$, in (ScZrNb)$_{1-x}$(RhPd)$_x$. For $x = 0.45$ no superconductivity above 1.7~K was observed~\cite{hea-cscl-cava}.} 
\end{figure}

The situation is expected to be different for the phonon spectrum, but
theoretical studies of phonons in random structures are very computationally demanding, and they were not undertaken for TNHZT.
For phonons in alloys, disorder induces an effect similar to that observed for the electronic structure: smearing of the dispersion relations, which is caused by fluctuations in the mass and the force constants \cite{phonon_broadening}.
A quantity which may help to predict whether phonons will be strongly scattered by disorder is
the mass-fluctuation phonon scattering parameter~\cite{phonon_broadening,Klemens_1955}:
\begin{equation}\label{eq:mass_fluct}
\Gamma_M = \sum_i \frac{c_i(M_i - \bar M)^2}{{\bar M}^2},
\end{equation}
where $c_i$ and $M_i$ are the concentration and atomic mass of the
$i$-th component and $\bar M$ is the average mass of the alloy. The phonon scattering rate (inverse of the
relaxation time) is proportional to $\Gamma_M$ \cite{Klemens_1955}.
Although it cannot fully cover variations in phonon broadenings with the alloy composition (see \cite{phonon_broadening} for discussion), in the absence of phonon calculations it can serve as an indicator of whether we can expect a strong disorder effects on phonons. The smaller the $\Gamma_M$, the weaker the disorder effect on the phonon spectrum may be expected, and the phonon structure of the alloy will be close to the spectrum predicted by the averaged mass and force constant model~\cite{phonon_broadening}.
In TNHZT, $\Gamma_M \simeq 0.15$, which is a moderate value, for which we may expect important modifications of the phonon spectrum due to the presence of disorder, but this subject has not yet been explored. 

As far as the superconductivity pairing mechanism in TNHZT is concerned, the experimental data~\cite{kozejl_2014} and theoretical calculations~\cite{jasiewicz_2016} imply a conventional electron-phonon superconductivity mechanism.
Moreover, superconductivity in TNHZT does not seem to be strongly affected by disorder.
The measured specific heat jump, $\Delta C/\gamma T_c = 1.63$~\cite{kozejl_2014}, is quite typical; other thermodynamic parameters in the superconducting phase also followed the conventional picture, with no special signatures of suppression of superconductivity by disorder. 
Theoretical calculations of the electron-phonon coupling (EPC) parameter $\lambda$ point towards the strong coupling limit, with the estimated $\lambda \sim 1$, in agreement with electronic specific heat analysis~\cite{jasiewicz_2016}.
In these calculations, the electronic contribution to the coupling constant, in the form of McMillan-Hopfield parameters, was computed using the KKR-CPA method, whereas the phonon contribution was estimated on the basis of the experimental measurements of the Debye temperature (see also the next section).
Because the experimental Debye temperature was used, the effect of disorder on the phonon structure was, in fact, taken into account in the calculations of $\lambda$. 
With these calculations, reproduction of the experimental value of $T_c$ using the calculated $\lambda$ required an enhanced value of the Coulomb pseudopotential parameter $\mu^* ~\sim 0.2$, which may indicate some suppressing effect of disorder on the critical temperature. However, since $\mu^*$ values of similar order were previously used to reproduce $T_c$ for a number of ordered structures, including elemental niobium, and the whole procedure involved approximations, this could not be used as an argument for concluding on the suppressing role of disorder on superconductivity in TNHZT.

\begin{table*}[t]
\caption{\label{concentrations} Lattice parameters $a$ (\AA) and atomic concentrations at the (1a) and (1b) sites in (ScZrNb)$_{1-x}$(RhPd)$_x$ for different $x$. 
Sc, Zr, and Nb1 atoms are on the (1a) (0,0,0) site, while Rh, Pd, and Nb2 are on the (1b) (0.5, 0.5, 0.5) site of the CsCl-type cubic structure (space group no. 221, Pm-3m). Data after~\cite{hea-cscl-cava}.}
\begin{ruledtabular}
\begin{tabular}{llllllll}
$x$ & $a$ (\AA) &  Sc &	Zr & Nb1 & Rh & Pd & Nb2\\
\hline
  0.35&3.293&0.433&0.433&0.133 & 0.35&0.35&0.30\\
  0.37&3.288&0.420&0.420&0.160&0.37&0.37&0.26\\
  0.40&3.281&0.400&0.400&0.200&0.40&0.40&0.20\\
  0.42&3.276&0.387&0.387&0.227&0.42&0.42&0.16\\
  0.45&3.268&0.367&0.367&0.267&0.45&0.45&0.10\\
\end{tabular}
\end{ruledtabular}
\end{table*}

A similar situation was found in a later reported TNHZT variant of Ta$_{0.335}$Nb$_{0.335}$Hf$_{0.11}$Zr$_{0.11}$Ti$_{0.11}$~\cite{tnhzt1,tnhzt2} which has a slightly higher $T_c = 7.7$~ K. 
For this alloys, superconductivity was also studied under extreme external pressure \cite{guo-pressure}, and $T_c$ was found to increase up to about 10~K and remained approximately constant to $\sim 100$ GPa. 
This trend in $T_c(p)$ also appeared to be consistent with the conventional electron-phonon picture.
Theoretical analysis~\cite{jasiewicz2019}, using a similar approach as for the first TNHZT variant  in~\cite{jasiewicz_2016},  showed a similar, small band smearing effects in its electronic structure.
By joining the calculated Mc\-Mil\-lan-Hop\-field parameters and measured Debye temperature, the calculations again predicted $\lambda \sim 1$ and reproduced the experimental $T_c$ with a similar and slightly enhanced Coulomb pseudopotential parameter value of $\mu^* \sim 0.2$.
Furthermore, with the help of the measured Gr\"uneisen parameter, the observed trend in $T_c(p)$ was correctly reproduced, as the product of several opposing tendencies. Although the EPC parameter $\lambda$ was found to decrease with pressure, the increase in the Debye temperature and the decrease in the Coulomb pseudopotential parameter $\mu^*$ resulted in an enhanced $T_c$.

\begin{figure}[b]
\centering
\includegraphics[width=0.95\columnwidth]{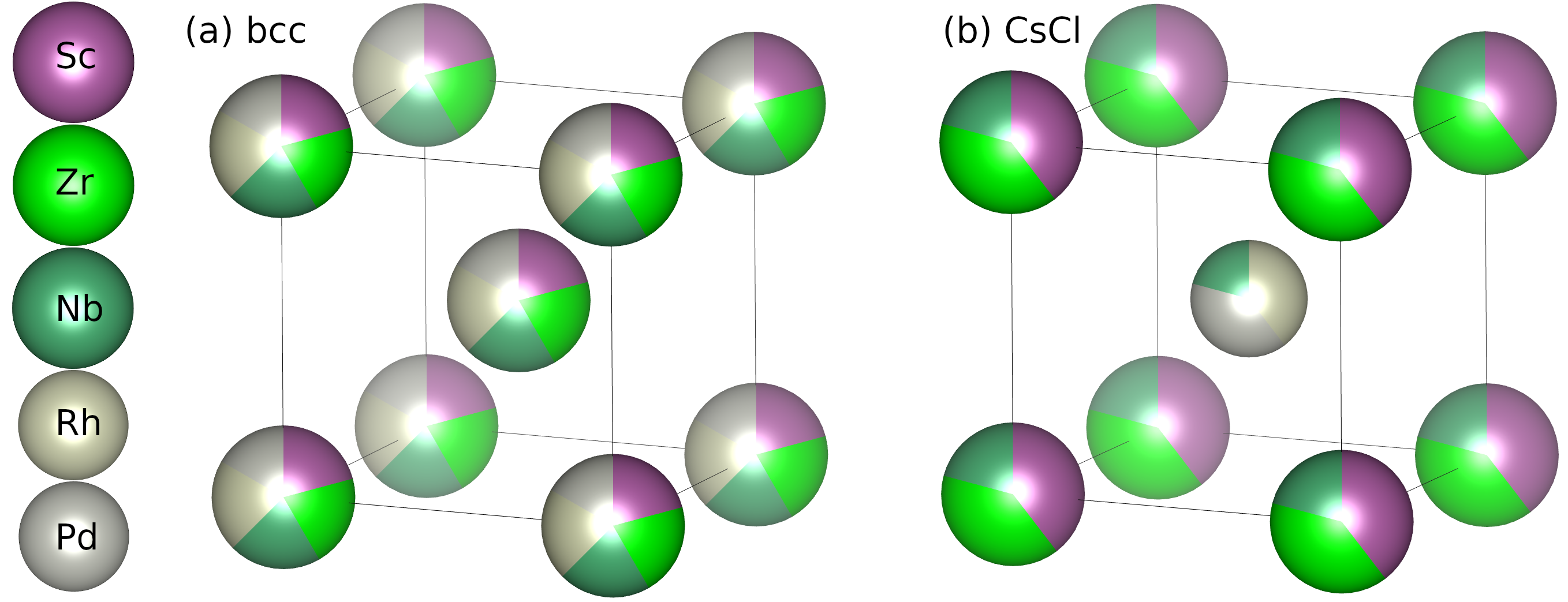}
\caption{\label{cscl_bcc}Comparison of the structures: a) simple ''monoatomic'' \textit{bcc} with all elements occupying the (equivalent) center and corner crystal sites; and b) CsCl, with two inequivalent sites and partial atomic ordering. Example drawn for $x = 0.40$, which has equal atomic concentrations of all elements in the unit cell, hence for the {\it bcc} variant occupation of all atoms at both sites is equal to 20\%, while for CsCl the corner (1a) position is occupied by Sc (40\%), Zr (40\%) and Nb1 (20\%), and the center (1b) position by Rh (40\%), Pd (40\%) and Nb2 (20\%).} 
\end{figure} 

Thus, previous works have shown that the superconductivity in TNHZT is conventional and well described by the electron-phonon coupling mechanism. The effect of disorder on the electronic structure is weak,
and we have some indications that superconductivity might be influenced by disorder but not to a large extent.

The impact of disorder on the electronic structure and superconductivity may be, of course, different in other HEAs, especially with a different crystal structure.
Thanks to experimental efforts, more superconducting HEAs were discovered, including a hexagonal variant  Re$_{0.56}$Nb$_{0.11}$Ti$_{0.11}$Zr$_{0.11}$Hf$_{0.11}$~\cite{marik2019_hex} and HEA-type compounds, that are partially ordered structures with HEA-like sublattices \cite{hea-sup1,hea-cscl-cava,stolze2018,mizuguchi2021,kasem2021}.
The latter materials have more complicated crystallographic structures of $\alpha-$Mn \cite{stolze2018}, CsCl \cite{hea-cscl-cava}, or recently discovered CuAl$_2$-type \cite{mizuguchi2021,kasem2021}.

In this work, we focus on (ScZrNb)$_{1-x}$(RhPd)$_x$ system (SZNRP in short), being the first example of a HEA-type compound.
For $x \in (0.35, 0.45)$ it was reported to form a CsCl-like structure \cite{hea-cscl-cava}.
Figure \ref{cscl_bcc} shows its unit cell and the relationship with the fully random {\it bcc} structure.
X-ray diffraction studies~\cite{hea-cscl-cava} showed that Sc and Zr atoms are located at the (1a) site, Rh and Pd at the (1b) site, while Nb atoms occupy both sublattices\footnote{The unit cell with exchanged (1a) and (1b) sites is equivalent, as it differs only by a shift in the origin of the coordinate system. Here we adopt the convention analogous to the CsCl structure, where the more electronegative element is at the (1b) site.}.
The experimental superconducting critical temperature~\cite{hea-cscl-cava} as a function of $x$ is presented in Fig.~\ref{tc_x}. 
The variation of $T_c$ with $x$ is remarkable. The highest critical temperature, of about 9.3~K, was found for $=0.35$, and decreases quite rapidly with increasing concentrations of Pd and Rh, dropping below 1.7~K (detection limit in \cite{hea-cscl-cava}) for $x = 0.45$. This trend is opposite to what can be expected on the basis of the Matheisen rule when analyzing the change in the number of valence electrons per atom (see~\cite{hea-cscl-cava} for more details).

The electronic structure of (ScZrNb)$_{1-x}$(RhPd)$_x$ is studied in this work. 
Densities of states, electronic dispersion relations, Fermi surfaces, and electronic contribution to the e\-le\-ctron-pho\-non coupling parameter (Mc\-Mil\-lan-Hop\-field parameters) are calculated as a function of alloy composition, $x$.
Quite surprisingly, we found that despite the partial ordering of the structure, the role of chemical disorder is more important here than in the previously mentioned case of a purely random {\it bcc}-type TNHZT system. 
The band structure appears to be strongly smeared as a result of enhanced electron scattering. 
Calculations of the mean-free electronic path $d$ show that in fact superconductivity in SZNRP appears on the border of the Mott-Ioffe-Regel limit \cite{ioffe1960non,hussey2004universality}, as the computed $d$ is of the order of the interatomic distance. This may have an influence on the thermodynamic parameters of the superconducting phase; however, most of them have not yet been reported.
Our studies show that superconductivity and electron-phonon coupling in SZNRP occurs to be more challenging to describe than in the TNHZT alloys, and that it offers a chance to investigate the interplay of superconductivity and the strong disorder effects.

\section{Methods}

Electronic structure calculations were performed using the Kor\-rin\-ga-Kohn-Rostoker method with the coherent potential approximation (KKR-CPA)~\cite{soven_1967,kaprzyk_1990,bansil_1999,stopa_2004} to account for the atomic disorder. The local density approximation (LDA) of Perdew and Wang~\cite{perdew_1992} was used to calculate the effective crystal potential in the spherical potential approximation and semi-relativistic treatment. The angular momentum cutoff $l_{max}$ = 3 was used. 
The radius of the muffin-tin (MT) spheres was set to the largest value for non-overlapping spheres, equal to $a\sqrt{3}/4$, where $a$ is the lattice parameter.  Fermi level ($E_F$) was accurately determined using the generalized Lloyd formula \cite{kaprzyk_1990}. 
The densities of states were computed on a fine mesh of 1540 points in the irreducible part of the Brillouin zone.
In addition to the electronic structure, we study the electronic contribution to the electron-phonon interaction parameter $\lambda$ by computing the McMillan-Hopfield parameters $\eta_i$ using the rigid muffin tin approximation (RMTA).
~\cite{gaspari_1972,gomersall_1974,papa_a15,mazin_1990,wiendlocha_2006,wiendlocha2008,wiendlocha2014}. 
In this approach, the electron-phonon interaction is decoupled into electronic and lattice contributions. The coupling parameter $\lambda$ is expressed as:
\begin{equation}\label{eq:lambda0}
\lambda = \sum_i \frac{c_i\eta_i}{M_i\langle{\omega_i^2}\rangle},
\end{equation}
where $\eta_i$ are calculated for each type of atom $i$ in the unit cell, $M_i$ is the atomic mass,  $\langle{\omega_i^2}\rangle$ is the average square atomic vibration frequency, and $c_i$ is the population of atoms in the unit cell, in the case of alloys $c_i$ becomes the atomic concentration of the element~\cite{kaprzyk_1996,jasiewicz2019}.
Within RMTA, the McMillan-Hopfield parameters are calculated as \cite{gaspari_1972,gomersall_1974,mazin_1990}:
\begin{equation}\label{eq:eta}
\eta_i =\!\sum_l \frac{(2l + 2)\,n_l(E_F)\,
n_{l+1}(E_F)}{2(2l+1)(2l+3)N(E_F)} \left|\int_0^{R_{\mathsf{MT}}}\!\!r^2
R_l\frac{dV}{dr}R_{l+1} \right|^2\!,
\end{equation}
where $l$ is the angular momentum number, $V(r)$ is the self-consistent potential at site $i$, $R_\mathsf{MT}$ is the radius of the $i$-th MT sphere, $R_l(r)$ is a normalized regular solution of the radial Schr\"odinger equation, $n_l(E_F)$ is the $l$--th partial DOS at the Fermi level $E_F$, and $N(E_F)$ is the total DOS per primitive cell.
For a more detailed discussion of RMTA see Refs.~\cite{mazin_1990,wiendlocha_2006} and references therein. 
As already mentioned, this approach has recently been successfully applied to analyze the electron-phonon interaction in TNHZT alloys \cite{jasiewicz_2016,jasiewicz2019}.

KKR-CPA and RMTA methods are used to
investigate the electronic structure and trends in the electronic contribution to electron-phonon coupling in
(ScZrNb)$_{1-x}$(RhPd)$_x$ series of alloys, in the range $0.35 \leq x \leq 0.45$.
The concentrations of atoms on each sublattice for a given $x$, as well as the lattice parameters, are presented in Table~\ref{concentrations}.
In the cases mentioned above of ''monoatomic'' HEAs (with one inequivalent crystal site) it was possible to approximate the average phonon frequency $\langle{\omega_i^2}\rangle$ assuming that the phonon spectrum follows the Debye model. 
In such a case, the experimental Debye temperature may be used to evaluate $\langle{\omega_i^2}\rangle$ and to successfully calculate $\lambda$, as theoretical phonon calculations are still very challenging for such highly disordered structures.
However, in the case of SZNRP it is not possible, since the heat capacity measurements have not been reported yet, nor this alloy is a ''monoatomic'' system where the Debye approximation is expected to be a reliable model of the phonon spectrum. 
Thus, we limit the discussion of electron-phonon coupling to analyze the trend with the change in the composition of the alloy based on the electronic contribution, described by the McMillan-Hopfield parameters $\eta_i$.

\section{Results and discussion}

\begin{figure}[t]
 \centering
\includegraphics[width=0.95\columnwidth]{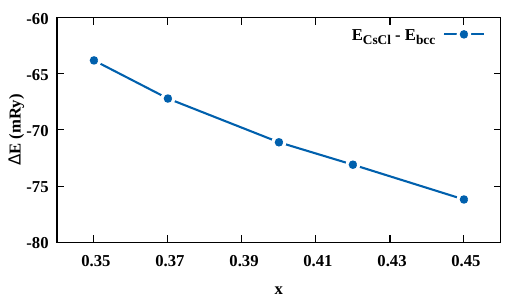}
\caption{\label{energy} The difference in the total energy of (ScZrNb)$_{1-x}$(RhPd)$_x$ alloys calculated in
the simple \textit{bcc} structure and the CsCl structure. The partially ordered CsCl-type structure has a considerably lower energy.} 
\end{figure}

\subsection{Structural preference}

\begin{figure*}[t!]
 \centering
\includegraphics[width=\textwidth]{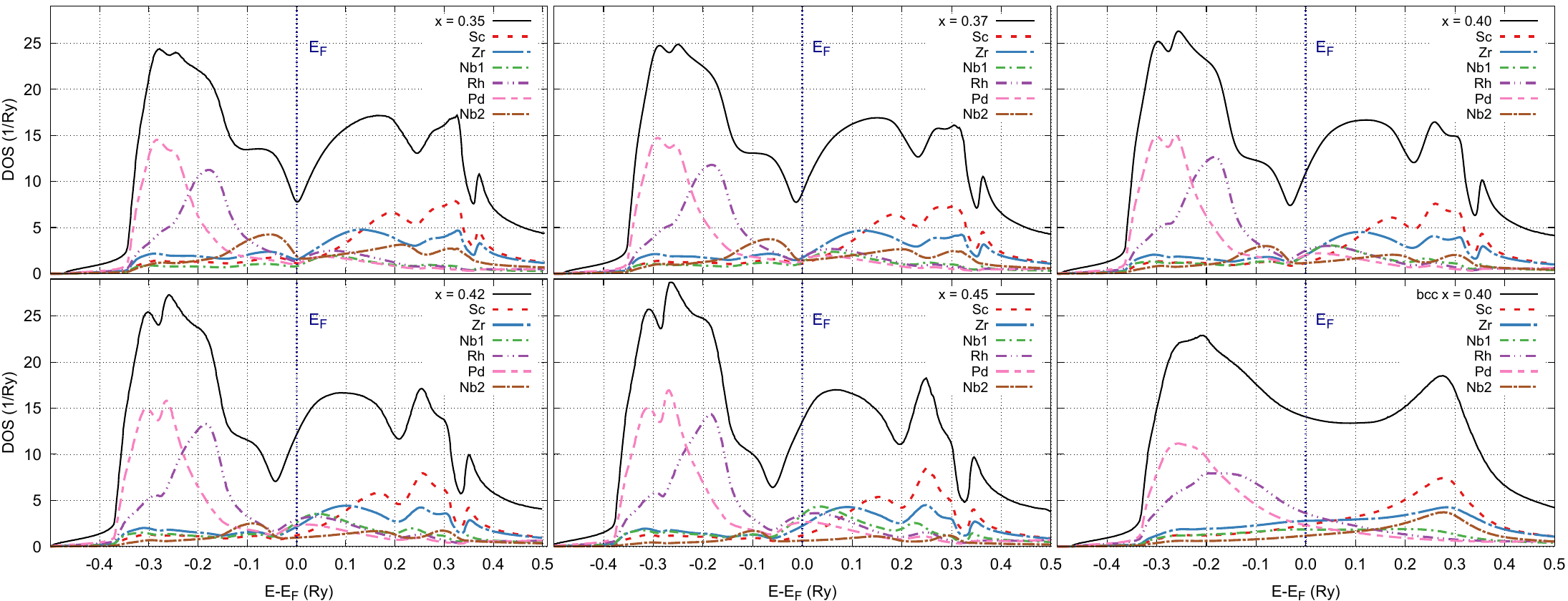}
\caption{\label{fig:tdos}Densities of states of (ScZrNb)$_{1-x}$(RhPd)$_x$. Total DOS per unit cell is plotted using a black continuous line; partial atomic densities (per atom) are plotted using dashed and dotted lines.} 
\end{figure*}

The first question which arises for the SZNRP system is whether we are able to explain why it forms in the partially ordered CsCl-like structure instead of a fully-disordered tung\-sten-type ''monoatomic'' \textit{bcc} structure.
To analyze this issue, we performed a series of calculations of the total energy for both the CsCl and the \textit{bcc} variants.
The graphical representation of these structures is shown in Fig.~\ref{cscl_bcc}. In the \textit{bcc} variant, the atoms are randomly distributed on a single site with probabilities according to their concentration. 
In the CsCl-type structure, two inequivalent crystal sites are occupied in a different way. According to the experiment, Sc and Zr are only located at (1a) (0.0, 0.0, 0.0) and Rh and Pd atoms only at (1b) (0.5,0.5,0.5), while Nb is distributed between both sites, with concentrations shown in Table~\ref{concentrations}. 

Figure~\ref{energy} shows the difference in the total energy of the unit cell, $\Delta E = E_{\rm CsCl} - E_{bcc}$, as a function of composition. For all $x$ the $\Delta E$ is negative; thus the energy of the partially ordered CsCl structure is lower than that of the fully random \textit{bcc} structure, and 
the simplest total energy criterion confirms the preference for partial ordering, observed in the (ScZrNb)$_{1-x}$(RhPd)$_x$ alloys. We analyze it further below, in terms of density of states and electronegativity of elements.

\subsection{Density of States}

The electronic densities of states (DOS) and their evolution with $x$ are plotted in Fig.~\ref{fig:tdos}.
In the last panel the DOS of the fully random variant \textit{bcc} is shown for $x = 0.40$.
On first glance, we see that in the CsCl structure a deep minimum of DOS is formed in the vicinity of the Fermi energy ($E_F$), in contrast to the \textit{bcc} case. 
This is in line with the structural preference and lower total energy of the partially ordered material, as the DOS shape in the fully random case leads to a higher electronic energy.

\begin{table*}[t]
\caption{\label{tab:dos} Total and partial atomic densities of states at the Fermi level in (1/Ry). Total DOS, $N(E_F)$, is given per unit cell, atomic densities of states are given per single atom.}
\begin{ruledtabular}
\begin{tabular}{llllllll}
$x$ & $N(E_F)$& Sc &	Zr & Nb1 & Rh & Pd & Nb2\\
\hline
0.35 & 16.50 & 4.26 & 5.85 & 9.86 & 6.28 & 4.84 & 10.34\\
0.37 & 18.57 & 4.27 & 6.69 & 12.96 & 7.87 & 6.36 & 9.01\\
0.40 & 21.38 & 4.78 & 8.20 & 16.90 & 10.21 & 8.19 & 9.42\\
0.42 & 23.62 & 5.05 & 8.95 & 18.82 & 11.25 & 8.86 & 9.79\\
0.45 & 26.25 & 5.38 & 9.94 & 21.02 & 12.30 & 9.50 & 10.25\\
\end{tabular}
\end{ruledtabular}
\end{table*}

\begin{figure}[b]
 \centering
\includegraphics[width=\columnwidth]{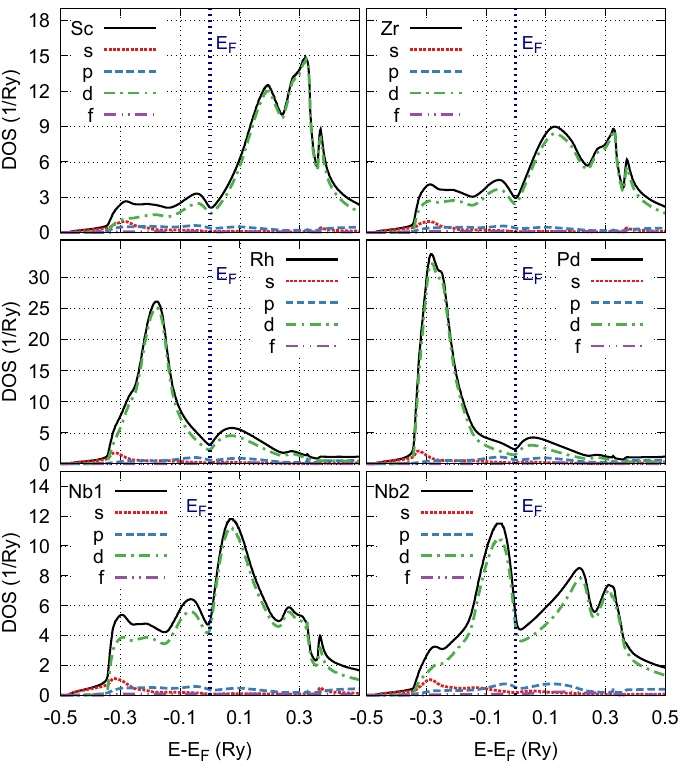}
\caption{\label{fig:dos35}Partial densities of states in SZNRP for $x=0.35$.} 
\end{figure}

We start the analysis of electronic densities of states with the $x = 0.35$ case, where $E_F$ is located at the bottom of the local minimum in DOS.
The partial atomic DOSes of this alloy are presented in Fig.~\ref{fig:dos35}. We can observe that the minimum in DOS is formed as a result of the separation of the $d$ states of the constituent atoms. The mostly unoccupied $3d$ shell of scandium and $4d$ of zirconium have their partial DOSes peaked above $E_F$, whereas the almost full $4d$ shells of rhodium and palladium develop DOS peaks in the lower energy range, about 0.25 Ry below $E_F$. 
As a consequence, the Fermi energy falls in between those peaks to the deep DOS minimum. 
It is interesting that the shape of the partial Nb DOS is different for the two inequivalent Nb atoms, located at the two different crystal sites. Nb1, which is at the (1a) position, has a DOS peak above $E_F$, that is, for unoccupied states, such as early transition metals Sc and Zr that are alloyed with Nb1 at the same site. On the other hand, Nb2, alloyed at the (1b) site with the late transition metals Rh and Pd, has a DOS peak below $E_F$, for the occupied states. This is associated with a larger filling of the d-shell of Nb2: inside the MT sphere, Nb2 has 0.5 more electrons than Nb1. 
These differences in partial DOSes and orbital occupation between the (1a) and (1b) sites are naturally correlated with the adopted CsCl crystal structure, and additionally explain why this system favors partial atom ordering.
The prototype structure, CsCl, is an ionic insulator, thus a material that is very different from the studied metallic alloy; nevertheless, it helps us to understand the mechanism of the partial ordering. 

\begin{figure}[b]
 \centering
\includegraphics[width=\columnwidth]{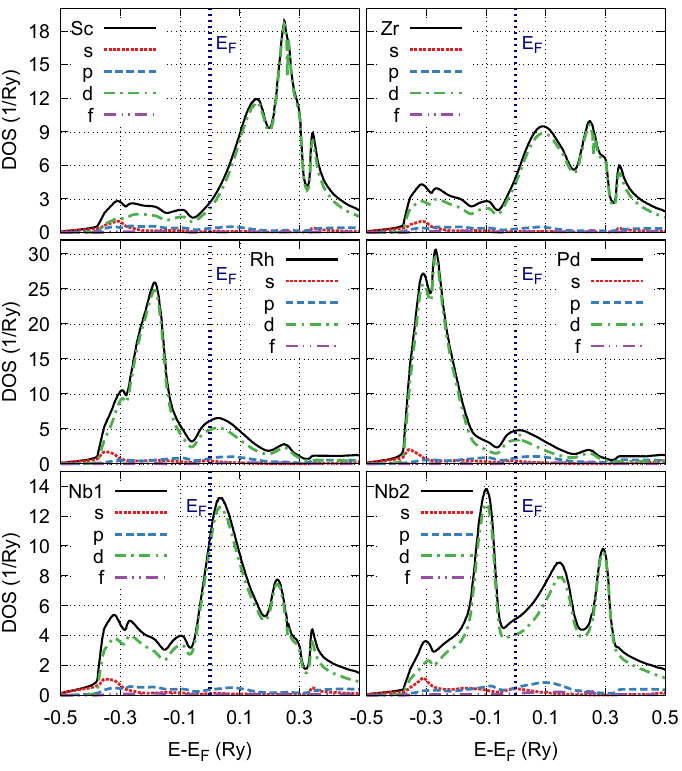}
\caption{\label{fig:dos45}Partial densities of states in SZNRP for $x=0.45$.} 
\end{figure}

In CsCl, the electropositive element Cs (Pauling electronegativity of 0.79) transfers its electron to the electronegative Cl (Pauling electronegativity of 3.16), stabilizing the structure through the ionic bonding. Similarly here, in SZNRP, the least electronegative metals Sc (1.36) and Zr (1.33) occupy position (1a), while the most electronegative metals Rh (2.28) and Pd (2.20) occupy position (1b). Nb, which is in between the two groups (electronegativity of 1.60), has no site preference and is located at both positions.
The transfer of electrons is in line with the
site preference: (1a) atoms donate more electrons to the electron gas in the interstitial region than (1b).
When comparing the total number of valence electrons inside the non-overlapping muffin tin spheres (which are of equal radius for both sites), Sc has 19.65 e$^-$ (1.35 e$^-$ less than the neutral Sc atom), Zr has 38.38 (1.62 e$^-$ less than the neutral Zr), Nb1: 39.46 (1.54 e$^-$ less than the neutral Nb) while Rh: 44.34 (0.66 e$^-$ less than the neutral Rh), Pd: 45.36 (0.64 e$^-$ less than the neutral Pd) and Nb2: 39.99 (1.01 e$^-$ less than the neutral Nb).
Computing the same quantities in the completely random \textit{bcc} variant, the numbers of electrons inside the MT spheres are around 0.2 e$^-$ higher for the Sc, Zr, and Nb1 atoms, and 0.2 e$^-$ lower for the Rh, Pd, and Nb2 atoms (Nb1 and Nb2 become equivalent), resulting in a less ionic but energetically unfavorable structure.
Thus, a mixed metallic-ionic bonding is responsible for the stability of the CsCl-type structure of SZNRP.

\begin{figure}[t]
 \centering
\includegraphics[width=0.95\columnwidth]{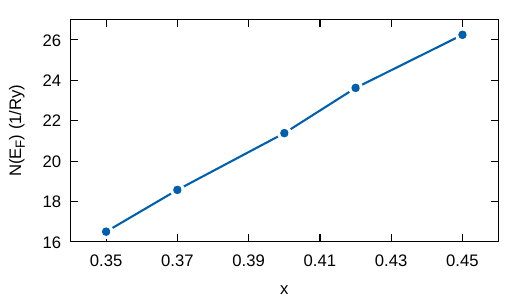}
\caption{\label{fig:ef} Total density of states at the Fermi level, $N(E_F)$, as a function of $x$ in (ScZrNb)$_{1-x}$(RhPd)$_x$.} 
\end{figure}

\begin{figure}[b]
 \centering
\includegraphics[width=0.95\columnwidth]{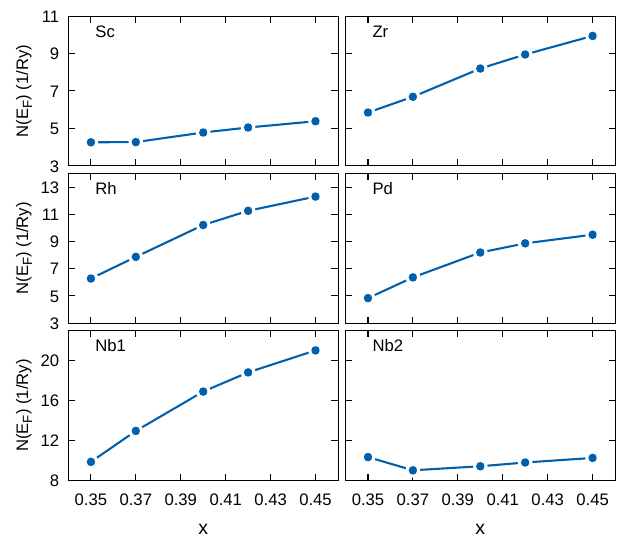}
\caption{\label{fig:pdos_ef}Partial atomic density of states at the Fermi level as a function of $x$ in (ScZrNb)$_{1-x}$(RhPd)$_x$.} 
\end{figure}

As far as charge transfers are concerned, when $x$ (that is, the concentration of Rh and Pd) increases, the (1b) site becomes more electronegative on average due to the lower content of Nb2, and the charge transfers from the (1a) atoms increase moderately (approximately 7\% for $x=0.45$). However, the DOS structure remains rigid, i.e. the primary effect of increasing $x$ is the shift of the Fermi level towards higher energy. This is due to the growing number of valence electrons. As a consequence, $E_F$ climbs the raising slope of DOS and the value of $N(E_F)$ increases, as shown in Fig.~\ref{fig:ef}.

The partial densities of the states for the border composition of $x = 0.45$ are plotted in Fig.~\ref{fig:dos45}. Moderate changes in the shapes of atomic DOSes can be observed, compared to $x = 0.35$. 
The movement of $E_F$ towards higher energy increases the DOS($E_F$) values for all atoms except Nb2, where $E_F$ moves from the decreasing slope of DOS through the minimum to increase again; see also Table~\ref{tab:dos} and Fig.~\ref{fig:pdos_ef}.
The total DOS at $E_F$, $N(E_F)$, as shown in Fig.~\ref{fig:ef}, increases linearly with $x$, reaching a 60\% higher value for $x = 0.45$. 
This is the first unexpected result obtained here, as this trend is opposite to the trend in the experimentally determined superconducting critical temperature $T_c$ (Fig.~\ref{tc_x}), which drops with $x$.

\subsection{McMillan-Hopfield Parameters}

\begin{figure}[htb]
\centering
  \includegraphics[width=\columnwidth]{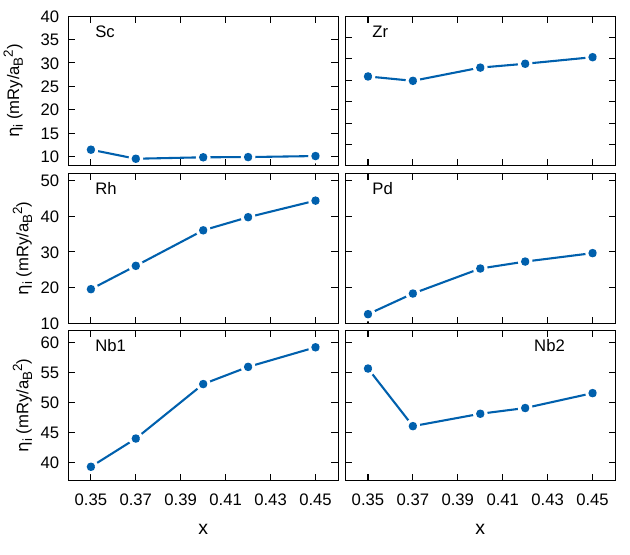}
  \caption{\label{fig:eta_all}McMillan-Hopfield parameters of elements for different $x$ parameters in (ScZrNb)$_{1-x}$(RhPd)$_x$.}
\end{figure}

From the computed band structure quantities, the McMil\-lan-Hop\-field parameters were calculated for each of the constituent atoms in the studied alloys. Values are collected in Table~\ref{tab:eta} and plotted as a function of $x$ in Fig.~\ref{fig:eta_all}. Scandium has the lowest $\eta_i$ on the order of 10 mRy/$a_B^2$, while both niobium atoms have the highest, between 40 and 60 mRy/$a_B^2$ ($a_B$ is the Bohr radius).
These values are lower compared to crystalline niobium (which is a {\it bcc} structure) where the McMillan-Hopfield parameter is about 76 mRy/$a_B^2$~\cite{jasiewicz_2016} when recalculated per equivalent unit cell.\footnote{Values of McMillan-Hopfield parameters have to be compared when calculated for the unit cells with equal number of atomic sites, values computed per primitive {\it bcc} cells with one atom are twice larger than the values computed in the cubic unit cell with two atoms, due to the two-time difference in the total density of states, $N(E_F)$, see eq.(\ref{eq:eta}). This difference is canceled when $\lambda$ is computed with eq.(\ref{eq:lambda0}), as the sum includes the number of atoms in the unit cell, giving the same final result for both elementary and primitive cells.}

\begin{table*}[t]
\caption{\label{tab:eta} McMillan-Hopfield parameters $\eta_i$ (mRy/a$_B^2$) for all the atoms, concentration-weighted sum of $\eta$ and average atomic mass for (1a) and (1b) sites.}
\begin{ruledtabular}
\begin{tabular}{lllllllllll}
$x$ & Sc &	Zr & Nb1 & Rh & Pd & Nb2 & $\sum_{(1a)} c_i\eta_i$& $\sum_{(1b)} c_i\eta_i$ &$\sum_{(1a)} c_iM_i$& $\sum_{(1b)} c_iM_i$\\
\hline
0.35&	11.45	&25.89	&39.27	&19.52	&12.52	&55.67&       	21.42&      	27.91& 71.40 & 101.14\\
0.37&	9.52	&24.87	&43.99	&26.08	&18.30	&46.05&       	21.48&      	28.39& 72.06 & 101.61\\
0.40&	9.82	&27.98	&53.09	&36.05	&25.31	&48.12&       	25.74&      	34.17& 73.05 & 102.31\\
0.42&	9.88	&28.87	&55.95	&39.73	&27.27	&49.08&       	27.66&      	35.99& 73.71 & 102.78\\
0.45&	10.10	&30.40	&59.22	&44.38	&29.64	&51.57&       	30.68&      	38.46& 74.79 & 103.49\\
\end{tabular}
\end{ruledtabular}
\end{table*}

\begin{figure}[b]
\centering
  \includegraphics[width=0.85\columnwidth]{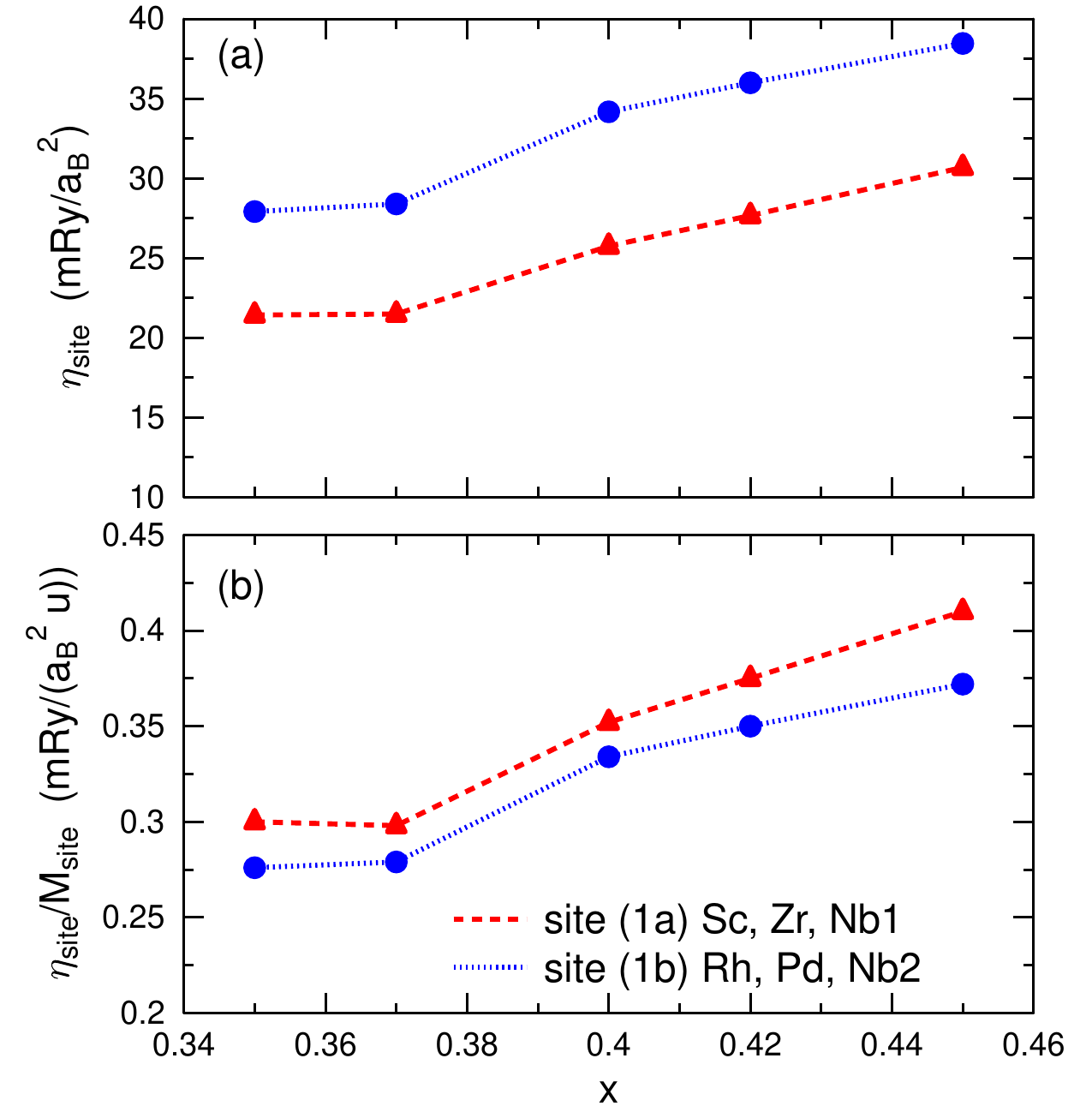}
  \caption{\label{fig:eta2}Panel (a): McMillan-Hopfield parameters summed over atoms on (1a) and (1b) sublattices, $\eta_{\rm site} = \sum_i'c_i\eta_i$, where $c_i$ is the concentration and sum is restricted to one sublattice. Panel (b): $\eta_{\rm site}$ divided by the average on-site atomic mass, $M_{\rm site} = \sum_i'c_iM_i$. Values plotted for different $x$ in (ScZrNb)$_{1-x}$(RhPd)$_x$.}
\end{figure}

When $x$ changes, $\eta_i(x)$ generally follow the trends in the densities of states and increase with $x$ for most atoms. 
For the Nb2 atom, where DOS($E_F$) decreases between 0.35 ad 0.37 and then increases, the same behavior is reflected in $\eta_{\rm Nb2}(x)$.
In Figure~\ref{fig:eta2}(a) we also plot the McMillan-Hopfield parameters summed over the crystal sites (1a) and (1b),
$\eta_{\rm site} = \sum_i'c_i\eta_i$ where $c_i$ is the atomic concentration and the sum is restricted to the selected sublattice. The numerical data are shown in Table~\ref{tab:eta}. 
The higher value is found for the (1a) site, and for both sites the sum increases with $x$.
The increasing tendencies of $\eta_i$ and $\eta_{\rm site}$ with $x$ are the second unexpected result obtained here, as similar to $N(E_F)$ it is not what we expect based on the experimental evolution of $T_c(x)$ shown in Figure~\ref{tc_x}. The decrease of the critical temperature suggests a decreasing value of the electron-phonon coupling parameter $\lambda(x)$, therefore we expected a decrease in the average $\eta(x)$.
McMillan-Hopfield parameters are of course only an electronic contribution to $\lambda$,
when it is computed using eq. (\ref{eq:lambda0}) $\eta_i$
are divided by the product $M_i\langle{\omega_i^2}\rangle$.
Due to the unknown phonon frequencies we cannot compute $\lambda$,
however, dividing $\eta_{\rm site}$ by the average mass of the atoms at the site, $M_{\rm site} = \sum_i'c_iM_i$, does not change the trend, as shown in
Figure~\ref{fig:eta2}(b). The $\eta_{\rm site}/M_{\rm site}$ ratio is higher for the (1a) site, and for both sites it continues to increase with $x$, despite the increase in the average mass of both sites.
A similar trend is maintained when, instead of dividing the site-averages of $\eta$ and $M$, the individual $\eta_i/M_i$ is first calculated and then summed with the appropriate concentrations. 
Furthermore, as the site-average masses increase with increasing $x$ (see Table~\ref{tab:eta}), we expect that the average phonon frequencies should decrease with $x$. In view of our results, $\lambda$ is rather expected to increase with $x$. 

\begin{figure*}[t]
\includegraphics[width=0.95\textwidth]{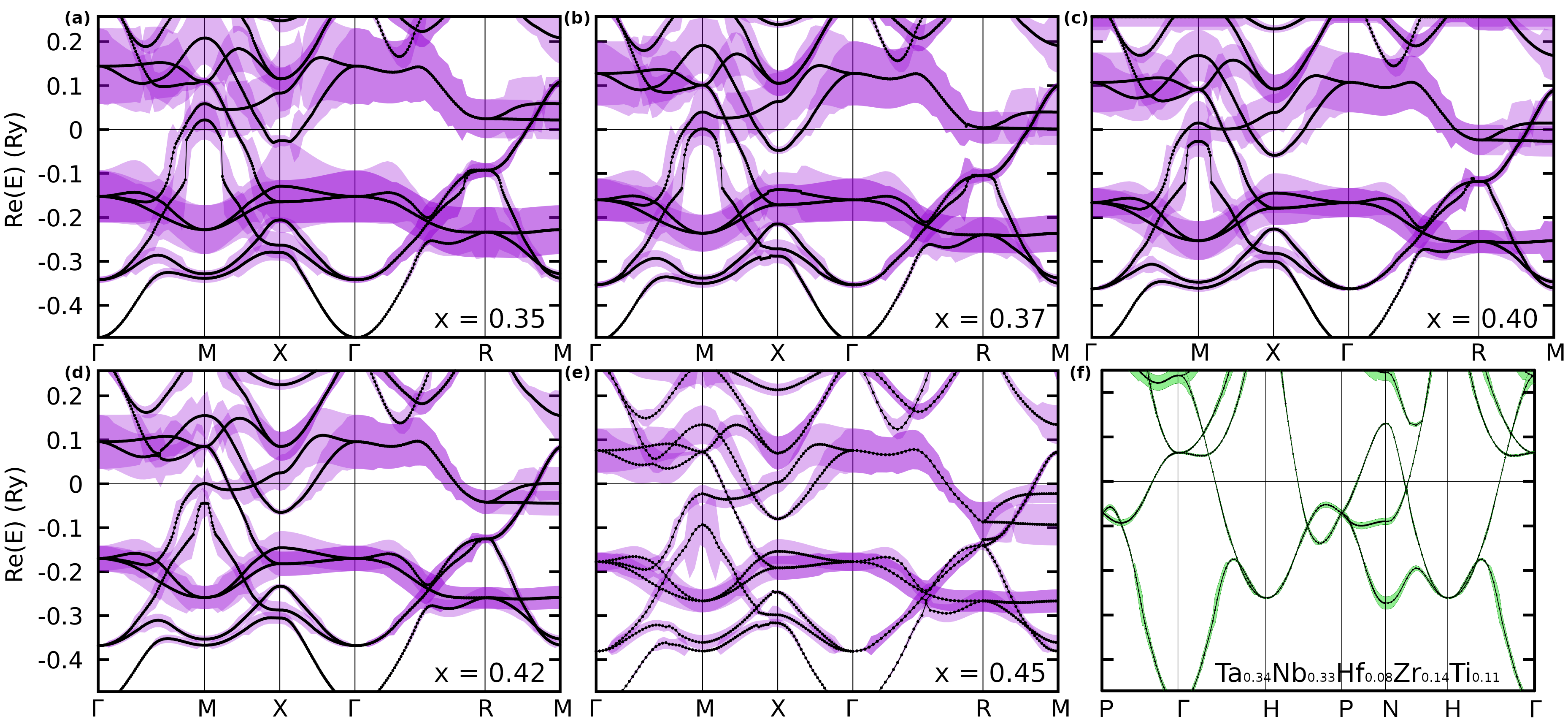}
\caption{\label{fig:bands}(a-e) Electronic band structure of the studied (ScZrNb)$_{1-x}$(RhPd)$_x$ alloys as a function of $x$. Black points are calculated real parts of the complex energy eigenvalues, whereas shading is a complex part describing the band smearing effect. The smearing is very strong in this series of alloys. 
For comparison in panel (f), complex bands are plotted for Ta$_{0.34}$Nb$_{0.33}$Hf$_{0.08}$Zr$_{0.14}$Ti$_{0.11}$, where Im$(E)$ is small and the bands near $E_F$ are very sharp.} 
\end{figure*}

Moreover, it is worth noting that the mass fluctuation parameter, defined in Eq. (\ref{eq:mass_fluct}), for SZNRP is not large and considerably smaller than in TNHZT. For $x=0.35$ it is equal to 0.10 for site (1a) and 0.003 for site (1b), for $x=0.45$ we have 0.09 (1a) and 0.0015 (1b), while in TNHZT it was about 0.15. 
One cannot exclude that disorder will have a strong influence on phonons, but at least the mass-fluctuation parameter does not suggest it.
Thus, either the rather unexpected strong increase in the average phonon frequencies $\langle{\omega_i^2}\rangle$ occurs along the (ScZrNb)$_{1-x}$(RhPd)$_x$ series or the drop in $T_c$ and no superconductivity above 1.7 K for $x = 0.45$ has a more complex origin, not resulting from the decrease in $\lambda$.
This could be the effect of a strong disorder manifesting itself in lowering the electronic lifetime (discussed in the next paragraph), the appearance of strong electronic correlations, or strong spin fluctuations. 
With increasing $x$ the concentration of palladium and rhodium increase. Both elements are known to exhibit strong spin fluctuations, which prevents Pd from being a superconductor \cite{berk,pd} and significantly reduces the $T_c$ of Rh to 0.3 mK \cite{rhodium-superconductivity,allen}. 
The possibility of enhancing spin fluctuations, or more generally, effective electron-electron interactions when increasing $x$ should be mentioned here, as it could explain the decrease in $T_c$ with $x$ via the increase of Coulomb pseudopotential $\mu^*$.
Heat capacity measurements and theoretical phonon calculations are desirable to explain the origin of the $T_c(x)$ trend, as the result of only electronic structure calculations cannot explain it.

\subsection{Electronic Bands and Fermi Surface}\label{sec:bands}

Calculations of the electronic bands within the KKR-CPA complex energy method allow to address the
fundamental question to what extent the chemical disorder influences the electronic properties of the material.
As we have already mentioned, disorder is a source of electron scattering, and the magnitude of this effect can be  studied by analyzing the band smearing and electronic lifetimes.

\begin{figure*}[t]
\includegraphics[width=0.85\textwidth]{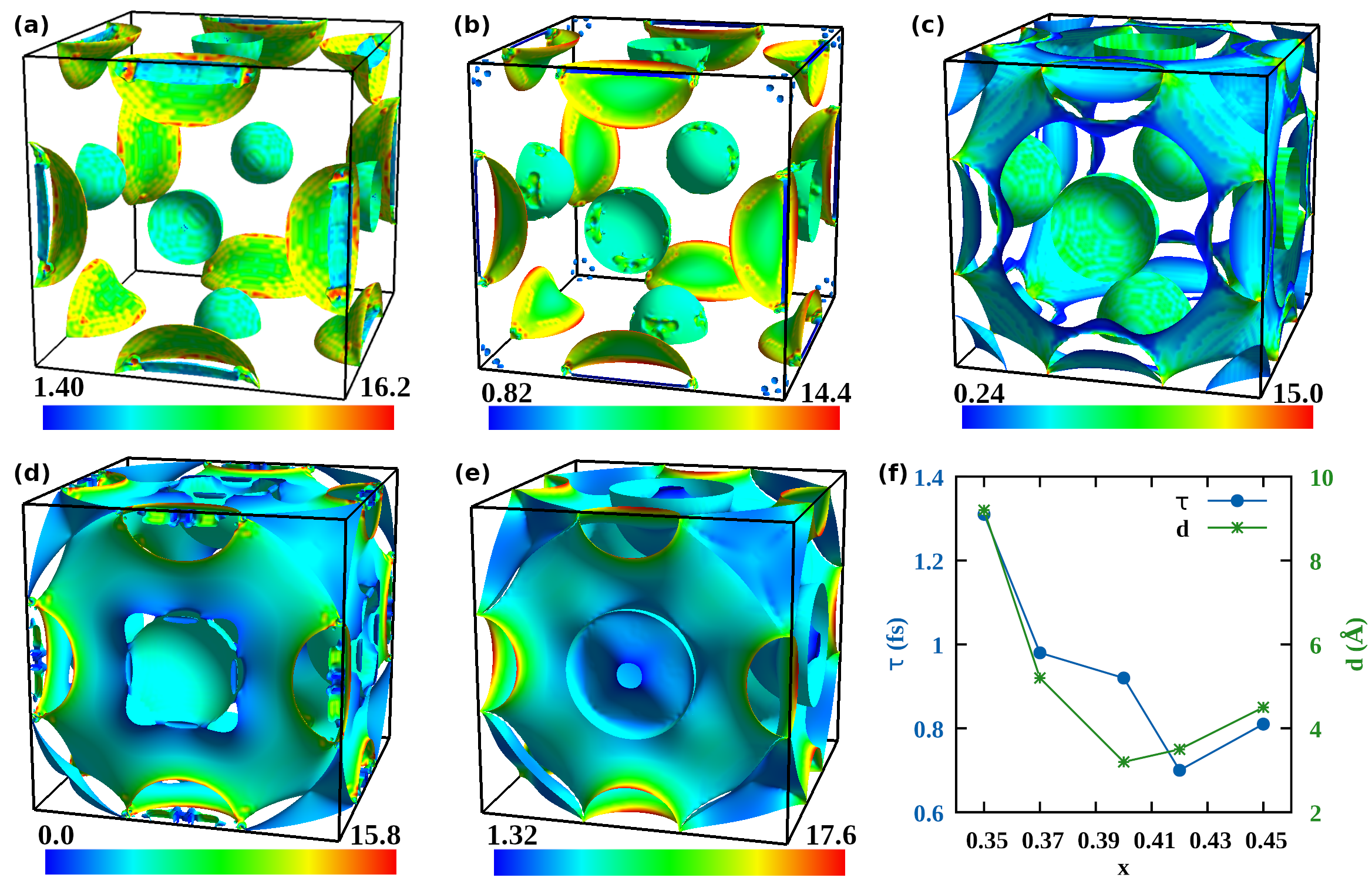}
\centering
\caption{\label{fig:fs}Fermi surface of the studied (ScZrNb)$_{1-x}$(RhPd)$_x$ alloys computed from the real part of the complex energy (representing the intersection of the center of the band with $E_F$), as a function of $x$, from $x = 0.35$ in panel (a) to $x = 0.45$ in panel (e). The color scale represents the electronic lifetime computed from the imaginary part of energy (Eq. \ref{eq:tau}) in femtoseconds (fs). Panel (f) shows the average $\tau$ in (fs) and  mean free path $d$ in (\AA).} 
\end{figure*}

The results obtained for the series of (ScZrNb)$_{1-x}$(RhPd)$_x$ alloys are shown in Figure~\ref{fig:bands}(a-e).
The real part of the energy describes the position of the center of the band (marked with black points), and the imaginary part describes the smearing of the band (violet shading).
The electronic lifetime $\tau$ is inversely proportional to the complex part of the energy, $\tau = \hbar/{2{\rm Im}(E)}$.
In each case, three bands cross the Fermi level, and thus SZNRP are multiband superconductors. 
What immediately catches the eye is the huge band smearing effect, which dominates the band pictures (the shading, corresponding to the imaginary part of energy, is not re-scaled and smearing reaches 0.1 Ry in large parts of the Brillouin zone).
For some k-points, it was even not possible to numerically find the center of the band, and in such cases the bands are interpolated (parts plotted with continues lines in Figure~\ref{fig:bands}).
For comparison, in panel (f), complex bands are shown for the TNHZT superconductor.
The striking difference between SZNRP and TNHZT is evident and may seem counterintuitive, since
the fully disordered TNHZT system has a much sharper bandstructure with a small band smearing when compared to the partially ordered structure of SZNRP.

The TNHZT alloy contains 5 elements that in the periodic table are located in neighboring groups $4^{\rm th}$ and 5$^{\rm th}$, therefore, are chemically and electronically more similar to each other than the elements of SZNRP. Here, each element is located in a different group of the periodic table, going from $3^{\rm rd}$ (Sc), through $4^{\rm th}$ (Zr), $5^{\rm th}$ (Nb) to $9^{\rm th}$ (Rh) and $10^{\rm th}$ (Pd), thus going from early transition metals with an open $d$ shell to late transition metals with a filled $d$ shell.
As mentioned above, the Pauling electronegativities of these elements cover a wide range from 1.33 (Zr) to 2.28 (Rh), in contrast to TNHZT, where they span a much smaller range between 1.3 (Hf) and 1.6 (Nb).
Figure~\ref{fig:bands} shows that even for the partially ordered case, such a mixture of very different electronic potentials leads to strong electron scattering, i.e., a strongly smeared band structure.
Therefore, the effect of disorder appears to be much more important for the determination of the electronic structure and properties of (ScZrNb)$_{1-x}$(RhPd)$_x$ than it was in the case of the Ta-Nb-Hf-Zr-Ti alloy.

In Figure \ref{fig:fs}(a-e), the Fermi surfaces (FS) of the series of SZNRP are presented. FSs were obtained based on the real part of the CPA complex band structure energy, thus they correspond to the band structure points that cross $E_F$ in Figure~\ref{fig:bands} and here the smearing effect of FS is not visualized. 
Fermi surfaces are colored with electronic lifetime $\tau_{j{\bf k}}$ defined by Eq.(\ref{eq:tau}) ($j$ numbers the bands). 
With increasing $x$ three effects are observed: the Fermi surface changes considerably its shape, the FS area becomes larger (in agreement with increasing $N(E_F)$), and the lifetime $\tau$ decreases.

The predicted strong electron scattering results in a very short electronic lifetime of the order of 1 fs.
{This is a much shorter value than in TNHZT, where our calculations predict $\tau = 89$ fs. Such a large difference of lifetimes is a consequence of the differences in the complex dispersion relations, discussed above and presented in Fig.~\ref{fig:bands}}.
The average lifetime (averaged over the Fermi surface) as a function of composition is shown in Fig. \ref{fig:fs}(f), together with the mean free path $d$, calculated as 
\begin{equation}
d_{j{\bf k}}=|{\bf v}_{j{\bf k}}|\tau_{j{\bf k}},
\end{equation}
and also averaged over FS. 
Fermi velocity is calculated as ${\bf v}_{j{\bf k}}=\nabla_{{\bf k}} \epsilon_{j{\bf k}}$ from the real part of the energy $\epsilon_{j{\bf k}}$ of electron in $j$-th band at ${\bf k}$.
Small values of $\tau$ lead to very small mean free paths $d$, of the order of single angstroms:
$d$ of all studied compositions varies between 3.2~\AA~and 9.2~\AA \ and is comparable to the smallest interatomic distance $d_{\rm min}$ which ranges from  2.83~\AA~to 2.85~\AA.
Similar values of mean free path (7 - 9~\AA) have been previously found in the case of Ni$_{0.25}$Fe$_{0.25}$Co$_{0.25}$Cr$_{0.25}$  alloy \cite{extreme_smearing_prl}, where extreme Fermi surface smearing was found.
This means that superconductivity in (ScZrNb)$_{1-x}$(RhPd)$_x$ compounds appear on the border of the Mott-Ioffe-Regel limit, for which the mean free path reaches interatomic distances and below which the quasiparticle picture of electrons in the periodic environment becomes invalid.
{In contrast, in TNHZT the much longer lifetime leads to a considerably larger mean-free path value of 459 \AA. This again highlights the difference between these two cases of superconducting high entropy alloys}.

As such strong electron scattering is predicted in SZNRP alloys and the tendency in $\tau(x)$ (and $d(x)$) generally follows the tendency in the superconducting critical temperature $T_c(x)$ (except for $x = 0.45$, where small increase in $\tau$ is observed), the natural question arises of whether the reduction in the critical temperature can be explained by the growing strength of electron scattering. For sure, the $T_c(x)$ and $\tau(x)$ trends appear to be correlated, in contrast to the trend observed for the density of states and McMillan-Hopfield parameters, which suggest that an increase in $T_c(x)$ should be observed. 
As we have already mentioned, heat capacity measurements and theoretical phonon calculations have to be performed to verify whether the trend in $T_c(x)$ can be explained conventionally (as due to the stiffening of the crystal lattice, which could reverse the increasing trend in $\eta(x)$ resulting in a decrease in $\lambda(x)$) or must be attributed to the effectively enhanced de-pairing interactions, usually described by the Coulomb repulsion parameter $\mu^*$. This can be the effect of strong disorder, electronic correlations, or spin fluctuations, all of which may explain the decrease in $T_c$~\cite{anderson_degradation,berk}.

\section{Summary}

Theoretical calculations of the electronic structure and McMillan-Hopfield parameters have been presented for the superconducting high-entropy-type system 
(ScZrNb)$_{1-x}$(RhPd)$_x$, with $0.35 \leq x \leq 0.45$. The partial ordering of the structure in the CsCl-type unit cell, reported in the experiment, is confirmed by the lower total energy of this structure, compared to the fully random {\it bcc}, and is correlated with the differences in Pauling electronegativites of the elements. 
The electronic densities of states are formed mainly by the d orbitals, and the total DOS at the Fermi level $N(E_F)$ increases linearly with $x$. The same increasing trend is observed for the McMillan-Hopfield parameters $\eta$ summed over all the atoms occupying each of the crystal sites. 
This tendency is opposite to the experimental results for the superconducting critical temperature, which was found to decrease with $x$. 
The electronic band structure has been analyzed using the complex energy method, and a very strong band smearing effect has been observed, much stronger than in the previously studied fully random Ta-Nb-Hf-Zr-Ti alloy. 
The calculated electronic mean free paths are of the order of single angstroms, which places the system on the border of the Mott-Ioffe-Regel limit.
This shows that despite the fact that SZNRP is a partially ordered structure, disorder is more important for the electronic structure of the (ScZrNb)$_{1-x}$(RhPd)$_x$ system than it was for the Ta-Nb-Hf-Zr-Ti alloy, although the latter is a completely random alloy. It is possible that the experimentally observed decreasing trend in the superconducting critical temperature $T_c(x)$ is a result of enhanced electron scattering due to disorder, as the electronic lifetime $\tau(x)$ has a similar decreasing trend with $x$ as the critical temperature, in contrast to the trends in $N(E_F)(x)$ and $\eta(x)$.

In conclusion, (ScZrNb)$_{1-x}$(RhPd)$_x$ high-entropy-type system is qualitatively different from the {\it bcc}-type TNHZT superconducting alloys and opens a unique opportunity to investigate the interplay of strong disorder and superconductivity.



\section*{Acknowledgements}
The work at AGH University was supported by the National Science Centre (Poland), project No. 
2017/26/E/ST3/00119.
We also gratefully acknowledge the Polish high-performance computing infrastructure PLGrid (HPC Centers: ACK Cyfronet AGH) for providing computer facilities and support within computational grant no. PLG/2022/015620.
\bibliography{refs-hea}

\begin{thebibliography}{54}%
\makeatletter
\providecommand \@ifxundefined [1]{%
 \@ifx{#1\undefined}
}%
\providecommand \@ifnum [1]{%
 \ifnum #1\expandafter \@firstoftwo
 \else \expandafter \@secondoftwo
 \fi
}%
\providecommand \@ifx [1]{%
 \ifx #1\expandafter \@firstoftwo
 \else \expandafter \@secondoftwo
 \fi
}%
\providecommand \natexlab [1]{#1}%
\providecommand \enquote  [1]{``#1''}%
\providecommand \bibnamefont  [1]{#1}%
\providecommand \bibfnamefont [1]{#1}%
\providecommand \citenamefont [1]{#1}%
\providecommand \href@noop [0]{\@secondoftwo}%
\providecommand \href [0]{\begingroup \@sanitize@url \@href}%
\providecommand \@href[1]{\@@startlink{#1}\@@href}%
\providecommand \@@href[1]{\endgroup#1\@@endlink}%
\providecommand \@sanitize@url [0]{\catcode `\\12\catcode `\$12\catcode
  `\&12\catcode `\#12\catcode `\^12\catcode `\_12\catcode `\%12\relax}%
\providecommand \@@startlink[1]{}%
\providecommand \@@endlink[0]{}%
\providecommand \url  [0]{\begingroup\@sanitize@url \@url }%
\providecommand \@url [1]{\endgroup\@href {#1}{\urlprefix }}%
\providecommand \urlprefix  [0]{URL }%
\providecommand \Eprint [0]{\href }%
\providecommand \doibase [0]{http://dx.doi.org/}%
\providecommand \selectlanguage [0]{\@gobble}%
\providecommand \bibinfo  [0]{\@secondoftwo}%
\providecommand \bibfield  [0]{\@secondoftwo}%
\providecommand \translation [1]{[#1]}%
\providecommand \BibitemOpen [0]{}%
\providecommand \bibitemStop [0]{}%
\providecommand \bibitemNoStop [0]{.\EOS\space}%
\providecommand \EOS [0]{\spacefactor3000\relax}%
\providecommand \BibitemShut  [1]{\csname bibitem#1\endcsname}%
\let\auto@bib@innerbib\@empty
\bibitem [{\citenamefont {Yeh}\ \emph {et~al.}(2004)\citenamefont {Yeh},
  \citenamefont {Chen}, \citenamefont {Lin}, \citenamefont {Gan}, \citenamefont
  {Chin}, \citenamefont {Shun}, \citenamefont {Tsau},\ and\ \citenamefont
  {Chang}}]{yeh_2004}%
  \BibitemOpen
  \bibfield  {author} {\bibinfo {author} {\bibfnamefont {J.-W.}\ \bibnamefont
  {Yeh}}, \bibinfo {author} {\bibfnamefont {S.-K.}\ \bibnamefont {Chen}},
  \bibinfo {author} {\bibfnamefont {S.-J.}\ \bibnamefont {Lin}}, \bibinfo
  {author} {\bibfnamefont {J.-Y.}\ \bibnamefont {Gan}}, \bibinfo {author}
  {\bibfnamefont {T.-S.}\ \bibnamefont {Chin}}, \bibinfo {author}
  {\bibfnamefont {T.-T.}\ \bibnamefont {Shun}}, \bibinfo {author}
  {\bibfnamefont {C.-H.}\ \bibnamefont {Tsau}}, \ and\ \bibinfo {author}
  {\bibfnamefont {S.-Y.}\ \bibnamefont {Chang}},\ }\bibfield  {title} {\enquote
  {\bibinfo {title} {{Nanostructured High-Entropy Alloys with Multiple
  Principal Elements: Novel Alloy Design Concepts and Outcomes}},}\ }\href
  {\doibase 10.1002/adem.200300567} {\bibfield  {journal} {\bibinfo  {journal}
  {Advanced Engineering Materials}\ }\textbf {\bibinfo {volume} {6}},\ \bibinfo
  {pages} {299--303} (\bibinfo {year} {2004})}\BibitemShut {NoStop}%
\bibitem [{\citenamefont {Yeh}\ \emph {et~al.}(2007)\citenamefont {Yeh},
  \citenamefont {Chen}, \citenamefont {Lin},\ and\ \citenamefont
  {Chen}}]{yeh_2007}%
  \BibitemOpen
  \bibfield  {author} {\bibinfo {author} {\bibfnamefont {Jien~Wei}\
  \bibnamefont {Yeh}}, \bibinfo {author} {\bibfnamefont {Yu~Liang}\
  \bibnamefont {Chen}}, \bibinfo {author} {\bibfnamefont {Su~Jien}\
  \bibnamefont {Lin}}, \ and\ \bibinfo {author} {\bibfnamefont {Swe~Kai}\
  \bibnamefont {Chen}},\ }\bibfield  {title} {\enquote {\bibinfo {title}
  {{High-Entropy Alloys - A New Era of Exploitation}},}\ }in\ \href {\doibase
  10.4028/www.scientific.net/MSF.560.1} {\emph {\bibinfo {booktitle} {Advanced
  Structural Materials III}}},\ \bibinfo {series} {Materials Science Forum},
  Vol.\ \bibinfo {volume} {560}\ (\bibinfo  {publisher} {Trans Tech
  Publications},\ \bibinfo {year} {2007})\ pp.\ \bibinfo {pages}
  {1--9}\BibitemShut {NoStop}%
\bibitem [{\citenamefont {Zhang}\ \emph {et~al.}(2014)\citenamefont {Zhang},
  \citenamefont {Zuo}, \citenamefont {Tang}, \citenamefont {Gao}, \citenamefont
  {Dahmen}, \citenamefont {Liaw},\ and\ \citenamefont {Lu}}]{ZHANG2014}%
  \BibitemOpen
  \bibfield  {author} {\bibinfo {author} {\bibfnamefont {Yong}\ \bibnamefont
  {Zhang}}, \bibinfo {author} {\bibfnamefont {Ting~Ting}\ \bibnamefont {Zuo}},
  \bibinfo {author} {\bibfnamefont {Zhi}\ \bibnamefont {Tang}}, \bibinfo
  {author} {\bibfnamefont {Michael~C.}\ \bibnamefont {Gao}}, \bibinfo {author}
  {\bibfnamefont {Karin~A.}\ \bibnamefont {Dahmen}}, \bibinfo {author}
  {\bibfnamefont {Peter~K.}\ \bibnamefont {Liaw}}, \ and\ \bibinfo {author}
  {\bibfnamefont {Zhao~Ping}\ \bibnamefont {Lu}},\ }\bibfield  {title}
  {\enquote {\bibinfo {title} {Microstructures and properties of high-entropy
  alloys},}\ }\href {\doibase https://doi.org/10.1016/j.pmatsci.2013.10.001}
  {\bibfield  {journal} {\bibinfo  {journal} {Progress in Materials Science}\
  }\textbf {\bibinfo {volume} {61}},\ \bibinfo {pages} {1--93} (\bibinfo {year}
  {2014})}\BibitemShut {NoStop}%
\bibitem [{\citenamefont {Tsai}\ and\ \citenamefont {Yeh}(2014)}]{tsai2014}%
  \BibitemOpen
  \bibfield  {author} {\bibinfo {author} {\bibfnamefont {Ming-Hung}\
  \bibnamefont {Tsai}}\ and\ \bibinfo {author} {\bibfnamefont {Jien-Wei}\
  \bibnamefont {Yeh}},\ }\bibfield  {title} {\enquote {\bibinfo {title}
  {High-entropy alloys: a critical review},}\ }\href {\doibase
  10.1080/21663831.2014.912690} {\bibfield  {journal} {\bibinfo  {journal}
  {Materials Research Letters}\ }\textbf {\bibinfo {volume} {2}},\ \bibinfo
  {pages} {107--123} (\bibinfo {year} {2014})}\BibitemShut {NoStop}%
\bibitem [{\citenamefont {Miracle}\ and\ \citenamefont
  {Senkov}(2017)}]{MIRACLE2017}%
  \BibitemOpen
  \bibfield  {author} {\bibinfo {author} {\bibfnamefont {D.B.}\ \bibnamefont
  {Miracle}}\ and\ \bibinfo {author} {\bibfnamefont {O.N.}\ \bibnamefont
  {Senkov}},\ }\bibfield  {title} {\enquote {\bibinfo {title} {A critical
  review of high entropy alloys and related concepts},}\ }\href {\doibase
  https://doi.org/10.1016/j.actamat.2016.08.081} {\bibfield  {journal}
  {\bibinfo  {journal} {Acta Materialia}\ }\textbf {\bibinfo {volume} {122}},\
  \bibinfo {pages} {448--511} (\bibinfo {year} {2017})}\BibitemShut {NoStop}%
\bibitem [{\citenamefont {George}\ \emph {et~al.}(2019)\citenamefont {George},
  \citenamefont {Raabe},\ and\ \citenamefont {Ritchie}}]{george2019}%
  \BibitemOpen
  \bibfield  {author} {\bibinfo {author} {\bibfnamefont {Easo~P}\ \bibnamefont
  {George}}, \bibinfo {author} {\bibfnamefont {Dierk}\ \bibnamefont {Raabe}}, \
  and\ \bibinfo {author} {\bibfnamefont {Robert~O}\ \bibnamefont {Ritchie}},\
  }\bibfield  {title} {\enquote {\bibinfo {title} {High-entropy alloys},}\
  }\href {\doibase 10.1038/s41578-019-0121-4} {\bibfield  {journal} {\bibinfo
  {journal} {Nature Reviews Materials}\ }\textbf {\bibinfo {volume} {4}},\
  \bibinfo {pages} {515--534} (\bibinfo {year} {2019})}\BibitemShut {NoStop}%
\bibitem [{\citenamefont {Murty}\ \emph {et~al.}(2019)\citenamefont {Murty},
  \citenamefont {Yeh}, \citenamefont {Ranganathan},\ and\ \citenamefont
  {Bhattacharjee}}]{murty2019}%
  \BibitemOpen
  \bibfield  {author} {\bibinfo {author} {\bibfnamefont
  {Bhagevatula~Satyanarayana}\ \bibnamefont {Murty}}, \bibinfo {author}
  {\bibfnamefont {Jien-Wei}\ \bibnamefont {Yeh}}, \bibinfo {author}
  {\bibfnamefont {Srinivasa}\ \bibnamefont {Ranganathan}}, \ and\ \bibinfo
  {author} {\bibfnamefont {PP}~\bibnamefont {Bhattacharjee}},\ }\href@noop {}
  {\emph {\bibinfo {title} {High-entropy alloys}}}\ (\bibinfo  {publisher}
  {Elsevier},\ \bibinfo {year} {2019})\BibitemShut {NoStop}%
\bibitem [{\citenamefont {Ko\ifmmode~\check{z}\else \v{z}\fi{}elj}\ \emph
  {et~al.}(2014)\citenamefont {Ko\ifmmode~\check{z}\else \v{z}\fi{}elj},
  \citenamefont {Vrtnik}, \citenamefont {Jelen}, \citenamefont {Jazbec},
  \citenamefont {Jagli\ifmmode \check{c}\else
  \v{c}\fi{}i\ifmmode~\acute{c}\else \'{c}\fi{}}, \citenamefont {Maiti},
  \citenamefont {Feuerbacher}, \citenamefont {Steurer},\ and\ \citenamefont
  {Dolin\ifmmode~\check{s}\else \v{s}\fi{}ek}}]{kozejl_2014}%
  \BibitemOpen
  \bibfield  {author} {\bibinfo {author} {\bibfnamefont {P.}~\bibnamefont
  {Ko\ifmmode~\check{z}\else \v{z}\fi{}elj}}, \bibinfo {author} {\bibfnamefont
  {S.}~\bibnamefont {Vrtnik}}, \bibinfo {author} {\bibfnamefont
  {A.}~\bibnamefont {Jelen}}, \bibinfo {author} {\bibfnamefont
  {S.}~\bibnamefont {Jazbec}}, \bibinfo {author} {\bibfnamefont
  {Z.}~\bibnamefont {Jagli\ifmmode \check{c}\else
  \v{c}\fi{}i\ifmmode~\acute{c}\else \'{c}\fi{}}}, \bibinfo {author}
  {\bibfnamefont {S.}~\bibnamefont {Maiti}}, \bibinfo {author} {\bibfnamefont
  {M.}~\bibnamefont {Feuerbacher}}, \bibinfo {author} {\bibfnamefont
  {W.}~\bibnamefont {Steurer}}, \ and\ \bibinfo {author} {\bibfnamefont
  {J.}~\bibnamefont {Dolin\ifmmode~\check{s}\else \v{s}\fi{}ek}},\ }\bibfield
  {title} {\enquote {\bibinfo {title} {{Discovery of a Superconducting
  High-Entropy Alloy}},}\ }\href {\doibase 10.1103/PhysRevLett.113.107001}
  {\bibfield  {journal} {\bibinfo  {journal} {Phys. Rev. Lett.}\ }\textbf
  {\bibinfo {volume} {113}},\ \bibinfo {pages} {107001} (\bibinfo {year}
  {2014})}\BibitemShut {NoStop}%
\bibitem [{\citenamefont {von Rohr}\ and\ \citenamefont {Cava}(2018)}]{tnhzt1}%
  \BibitemOpen
  \bibfield  {author} {\bibinfo {author} {\bibfnamefont {Fabian~O.}\
  \bibnamefont {von Rohr}}\ and\ \bibinfo {author} {\bibfnamefont {Robert~J.}\
  \bibnamefont {Cava}},\ }\bibfield  {title} {\enquote {\bibinfo {title}
  {{Isoelectronic substitutions and aluminium alloying in the Ta-Nb-Hf-Zr-Ti
  high-entropy alloy superconductor}},}\ }\href {\doibase
  10.1103/PhysRevMaterials.2.034801} {\bibfield  {journal} {\bibinfo  {journal}
  {Phys. Rev. Materials}\ }\textbf {\bibinfo {volume} {2}},\ \bibinfo {pages}
  {034801} (\bibinfo {year} {2018})}\BibitemShut {NoStop}%
\bibitem [{\citenamefont {von Rohr}\ \emph {et~al.}(2016)\citenamefont {von
  Rohr}, \citenamefont {Winiarski}, \citenamefont {Tao}, \citenamefont
  {Klimczuk},\ and\ \citenamefont {Cava}}]{tnhzt2}%
  \BibitemOpen
  \bibfield  {author} {\bibinfo {author} {\bibfnamefont {Fabian}\ \bibnamefont
  {von Rohr}}, \bibinfo {author} {\bibfnamefont {Micha{\l}~J.}\ \bibnamefont
  {Winiarski}}, \bibinfo {author} {\bibfnamefont {Jing}\ \bibnamefont {Tao}},
  \bibinfo {author} {\bibfnamefont {Tomasz}\ \bibnamefont {Klimczuk}}, \ and\
  \bibinfo {author} {\bibfnamefont {Robert~Joseph}\ \bibnamefont {Cava}},\
  }\bibfield  {title} {\enquote {\bibinfo {title} {{Effect of electron count
  and chemical complexity in the Ta-Nb-Hf-Zr-Ti high-entropy alloy
  superconductor}},}\ }\href {\doibase 10.1073/pnas.1615926113} {\bibfield
  {journal} {\bibinfo  {journal} {Proceedings of the National Academy of
  Sciences}\ }\textbf {\bibinfo {volume} {113}},\ \bibinfo {pages}
  {E7144--E7150} (\bibinfo {year} {2016})}\BibitemShut {NoStop}%
\bibitem [{\citenamefont {Sogabe}\ \emph {et~al.}(2018)\citenamefont {Sogabe},
  \citenamefont {Goto},\ and\ \citenamefont {Mizuguchi}}]{hea-sup1}%
  \BibitemOpen
  \bibfield  {author} {\bibinfo {author} {\bibfnamefont {Ryota}\ \bibnamefont
  {Sogabe}}, \bibinfo {author} {\bibfnamefont {Yosuke}\ \bibnamefont {Goto}}, \
  and\ \bibinfo {author} {\bibfnamefont {Yoshikazu}\ \bibnamefont
  {Mizuguchi}},\ }\bibfield  {title} {\enquote {\bibinfo {title}
  {{Superconductivity in {REO}0.5F0.5BiS2 with high-entropy-alloy-type blocking
  layers}},}\ }\href {\doibase 10.7567/apex.11.053102} {\bibfield  {journal}
  {\bibinfo  {journal} {Applied Physics Express}\ }\textbf {\bibinfo {volume}
  {11}},\ \bibinfo {pages} {053102} (\bibinfo {year} {2018})}\BibitemShut
  {NoStop}%
\bibitem [{\citenamefont {Sobota}\ \emph {et~al.}(2022)\citenamefont {Sobota},
  \citenamefont {Topolnicki}, \citenamefont {Ossowski}, \citenamefont {Pikula},
  \citenamefont {Pikul},\ and\ \citenamefont {Idczak}}]{prb_pikul}%
  \BibitemOpen
  \bibfield  {author} {\bibinfo {author} {\bibfnamefont {P.}~\bibnamefont
  {Sobota}}, \bibinfo {author} {\bibfnamefont {R.}~\bibnamefont {Topolnicki}},
  \bibinfo {author} {\bibfnamefont {T.}~\bibnamefont {Ossowski}}, \bibinfo
  {author} {\bibfnamefont {T.}~\bibnamefont {Pikula}}, \bibinfo {author}
  {\bibfnamefont {A.}~\bibnamefont {Pikul}}, \ and\ \bibinfo {author}
  {\bibfnamefont {R.}~\bibnamefont {Idczak}},\ }\bibfield  {title} {\enquote
  {\bibinfo {title} {Superconductivity in the high-entropy alloy
  ${(\mathrm{NbTa})}_{0.67}{(\mathrm{MoHfW})}_{0.33}$},}\ }\href {\doibase
  10.1103/PhysRevB.106.184512} {\bibfield  {journal} {\bibinfo  {journal}
  {Phys. Rev. B}\ }\textbf {\bibinfo {volume} {106}},\ \bibinfo {pages}
  {184512} (\bibinfo {year} {2022})}\BibitemShut {NoStop}%
\bibitem [{\citenamefont {Matthias}(1955)}]{matthias}%
  \BibitemOpen
  \bibfield  {author} {\bibinfo {author} {\bibfnamefont {B.~T.}\ \bibnamefont
  {Matthias}},\ }\bibfield  {title} {\enquote {\bibinfo {title} {Empirical
  relation between superconductivity and the number of valence electrons per
  atom},}\ }\href {\doibase 10.1103/PhysRev.97.74} {\bibfield  {journal}
  {\bibinfo  {journal} {Phys. Rev.}\ }\textbf {\bibinfo {volume} {97}},\
  \bibinfo {pages} {74--76} (\bibinfo {year} {1955})}\BibitemShut {NoStop}%
\bibitem [{\citenamefont {Stolze}\ \emph
  {et~al.}(2018{\natexlab{a}})\citenamefont {Stolze}, \citenamefont {Cevallos},
  \citenamefont {Kong},\ and\ \citenamefont {Cava}}]{stolze2018}%
  \BibitemOpen
  \bibfield  {author} {\bibinfo {author} {\bibfnamefont {Karoline}\
  \bibnamefont {Stolze}}, \bibinfo {author} {\bibfnamefont {F.~Alex}\
  \bibnamefont {Cevallos}}, \bibinfo {author} {\bibfnamefont {Tai}\
  \bibnamefont {Kong}}, \ and\ \bibinfo {author} {\bibfnamefont {Robert~J.}\
  \bibnamefont {Cava}},\ }\bibfield  {title} {\enquote {\bibinfo {title}
  {{High-entropy alloy superconductors on an $\sigma$-Mn lattice}},}\ }\href
  {\doibase 10.1039/C8TC03337D} {\bibfield  {journal} {\bibinfo  {journal} {J.
  Mater. Chem. C}\ }\textbf {\bibinfo {volume} {6}},\ \bibinfo {pages}
  {10441--10449} (\bibinfo {year} {2018}{\natexlab{a}})}\BibitemShut {NoStop}%
\bibitem [{\citenamefont {Stolze}\ \emph
  {et~al.}(2018{\natexlab{b}})\citenamefont {Stolze}, \citenamefont {Tao},
  \citenamefont {von Rohr}, \citenamefont {Kong},\ and\ \citenamefont
  {Cava}}]{hea-cscl-cava}%
  \BibitemOpen
  \bibfield  {author} {\bibinfo {author} {\bibfnamefont {Karoline}\
  \bibnamefont {Stolze}}, \bibinfo {author} {\bibfnamefont {Jing}\ \bibnamefont
  {Tao}}, \bibinfo {author} {\bibfnamefont {Fabian~O.}\ \bibnamefont {von
  Rohr}}, \bibinfo {author} {\bibfnamefont {Tai}\ \bibnamefont {Kong}}, \ and\
  \bibinfo {author} {\bibfnamefont {Robert~J.}\ \bibnamefont {Cava}},\
  }\bibfield  {title} {\enquote {\bibinfo {title} {{Sc–Zr–Nb–Rh–Pd and
  Sc–Zr–Nb–Ta–Rh–Pd High-Entropy Alloy Superconductors on a CsCl-Type
  Lattice}},}\ }\href {\doibase 10.1021/acs.chemmater.7b04578} {\bibfield
  {journal} {\bibinfo  {journal} {Chemistry of Materials}\ }\textbf {\bibinfo
  {volume} {30}},\ \bibinfo {pages} {906--914} (\bibinfo {year}
  {2018}{\natexlab{b}})},\ \Eprint
  {http://arxiv.org/abs/https://doi.org/10.1021/acs.chemmater.7b04578}
  {https://doi.org/10.1021/acs.chemmater.7b04578} \BibitemShut {NoStop}%
\bibitem [{\citenamefont {Anderson}(1959)}]{anderson_59}%
  \BibitemOpen
  \bibfield  {author} {\bibinfo {author} {\bibfnamefont {P.W.}\ \bibnamefont
  {Anderson}},\ }\bibfield  {title} {\enquote {\bibinfo {title} {Theory of
  dirty superconductors},}\ }\href {\doibase
  https://doi.org/10.1016/0022-3697(59)90036-8} {\bibfield  {journal} {\bibinfo
   {journal} {Journal of Physics and Chemistry of Solids}\ }\textbf {\bibinfo
  {volume} {11}},\ \bibinfo {pages} {26--30} (\bibinfo {year}
  {1959})}\BibitemShut {NoStop}%
\bibitem [{\citenamefont {Mackenzie}\ \emph {et~al.}(1998)\citenamefont
  {Mackenzie}, \citenamefont {Haselwimmer}, \citenamefont {Tyler},
  \citenamefont {Lonzarich}, \citenamefont {Mori}, \citenamefont {Nishizaki},\
  and\ \citenamefont {Maeno}}]{sr2ruo4-disorder}%
  \BibitemOpen
  \bibfield  {author} {\bibinfo {author} {\bibfnamefont {A.~P.}\ \bibnamefont
  {Mackenzie}}, \bibinfo {author} {\bibfnamefont {R.~K.~W.}\ \bibnamefont
  {Haselwimmer}}, \bibinfo {author} {\bibfnamefont {A.~W.}\ \bibnamefont
  {Tyler}}, \bibinfo {author} {\bibfnamefont {G.~G.}\ \bibnamefont
  {Lonzarich}}, \bibinfo {author} {\bibfnamefont {Y.}~\bibnamefont {Mori}},
  \bibinfo {author} {\bibfnamefont {S.}~\bibnamefont {Nishizaki}}, \ and\
  \bibinfo {author} {\bibfnamefont {Y.}~\bibnamefont {Maeno}},\ }\bibfield
  {title} {\enquote {\bibinfo {title} {{Extremely Strong Dependence of
  Superconductivity on Disorder in ${\mathrm{Sr}}_{2}{\mathrm{RuO}}_{4}$}},}\
  }\href {\doibase 10.1103/PhysRevLett.80.161} {\bibfield  {journal} {\bibinfo
  {journal} {Phys. Rev. Lett.}\ }\textbf {\bibinfo {volume} {80}},\ \bibinfo
  {pages} {161--164} (\bibinfo {year} {1998})}\BibitemShut {NoStop}%
\bibitem [{\citenamefont {Leroux}\ \emph {et~al.}(2019)\citenamefont {Leroux},
  \citenamefont {Mishra}, \citenamefont {Ruff}, \citenamefont {Claus},
  \citenamefont {Smylie}, \citenamefont {Opagiste}, \citenamefont {Rodière},
  \citenamefont {Kayani}, \citenamefont {Gu}, \citenamefont {Tranquada},
  \citenamefont {Kwok}, \citenamefont {Islam},\ and\ \citenamefont
  {Welp}}]{disorder_raises_cuprate}%
  \BibitemOpen
  \bibfield  {author} {\bibinfo {author} {\bibfnamefont {Maxime}\ \bibnamefont
  {Leroux}}, \bibinfo {author} {\bibfnamefont {Vivek}\ \bibnamefont {Mishra}},
  \bibinfo {author} {\bibfnamefont {Jacob P.~C.}\ \bibnamefont {Ruff}},
  \bibinfo {author} {\bibfnamefont {Helmut}\ \bibnamefont {Claus}}, \bibinfo
  {author} {\bibfnamefont {Matthew~P.}\ \bibnamefont {Smylie}}, \bibinfo
  {author} {\bibfnamefont {Christine}\ \bibnamefont {Opagiste}}, \bibinfo
  {author} {\bibfnamefont {Pierre}\ \bibnamefont {Rodière}}, \bibinfo {author}
  {\bibfnamefont {Asghar}\ \bibnamefont {Kayani}}, \bibinfo {author}
  {\bibfnamefont {G.~D.}\ \bibnamefont {Gu}}, \bibinfo {author} {\bibfnamefont
  {John~M.}\ \bibnamefont {Tranquada}}, \bibinfo {author} {\bibfnamefont
  {Wai-Kwong}\ \bibnamefont {Kwok}}, \bibinfo {author} {\bibfnamefont
  {Zahirul}\ \bibnamefont {Islam}}, \ and\ \bibinfo {author} {\bibfnamefont
  {Ulrich}\ \bibnamefont {Welp}},\ }\bibfield  {title} {\enquote {\bibinfo
  {title} {Disorder raises the critical temperature of a cuprate
  superconductor},}\ }\href {\doibase 10.1073/pnas.1817134116} {\bibfield
  {journal} {\bibinfo  {journal} {Proceedings of the National Academy of
  Sciences}\ }\textbf {\bibinfo {volume} {116}},\ \bibinfo {pages}
  {10691--10697} (\bibinfo {year} {2019})},\ \Eprint
  {http://arxiv.org/abs/https://www.pnas.org/doi/pdf/10.1073/pnas.1817134116}
  {https://www.pnas.org/doi/pdf/10.1073/pnas.1817134116} \BibitemShut {NoStop}%
\bibitem [{\citenamefont {Zhao}\ \emph {et~al.}(2019)\citenamefont {Zhao},
  \citenamefont {Lin}, \citenamefont {Xiao}, \citenamefont {Huang},
  \citenamefont {Yao}, \citenamefont {Yan}, \citenamefont {Xing}, \citenamefont
  {Zhang}, \citenamefont {Li}, \citenamefont {Hoshino} \emph
  {et~al.}}]{raises_nbse2}%
  \BibitemOpen
  \bibfield  {author} {\bibinfo {author} {\bibfnamefont {Kun}\ \bibnamefont
  {Zhao}}, \bibinfo {author} {\bibfnamefont {Haicheng}\ \bibnamefont {Lin}},
  \bibinfo {author} {\bibfnamefont {Xiao}\ \bibnamefont {Xiao}}, \bibinfo
  {author} {\bibfnamefont {Wantong}\ \bibnamefont {Huang}}, \bibinfo {author}
  {\bibfnamefont {Wei}\ \bibnamefont {Yao}}, \bibinfo {author} {\bibfnamefont
  {Mingzhe}\ \bibnamefont {Yan}}, \bibinfo {author} {\bibfnamefont {Ying}\
  \bibnamefont {Xing}}, \bibinfo {author} {\bibfnamefont {Qinghua}\
  \bibnamefont {Zhang}}, \bibinfo {author} {\bibfnamefont {Zi-Xiang}\
  \bibnamefont {Li}}, \bibinfo {author} {\bibfnamefont {Shintaro}\ \bibnamefont
  {Hoshino}},  \emph {et~al.},\ }\bibfield  {title} {\enquote {\bibinfo {title}
  {Disorder-induced multifractal superconductivity in monolayer niobium
  dichalcogenides},}\ }\href {\doibase 10.1038/s41567-019-0570-0} {\bibfield
  {journal} {\bibinfo  {journal} {Nature Physics}\ }\textbf {\bibinfo {volume}
  {15}},\ \bibinfo {pages} {904--910} (\bibinfo {year} {2019})}\BibitemShut
  {NoStop}%
\bibitem [{\citenamefont {Anderson}\ \emph {et~al.}(1983)\citenamefont
  {Anderson}, \citenamefont {Muttalib},\ and\ \citenamefont
  {Ramakrishnan}}]{anderson_degradation}%
  \BibitemOpen
  \bibfield  {author} {\bibinfo {author} {\bibfnamefont {P.~W.}\ \bibnamefont
  {Anderson}}, \bibinfo {author} {\bibfnamefont {K.~A.}\ \bibnamefont
  {Muttalib}}, \ and\ \bibinfo {author} {\bibfnamefont {T.~V.}\ \bibnamefont
  {Ramakrishnan}},\ }\bibfield  {title} {\enquote {\bibinfo {title} {Theory of
  the "universal" degradation of ${T}_{c}$ in high-temperature
  superconductors},}\ }\href {\doibase 10.1103/PhysRevB.28.117} {\bibfield
  {journal} {\bibinfo  {journal} {Phys. Rev. B}\ }\textbf {\bibinfo {volume}
  {28}},\ \bibinfo {pages} {117--120} (\bibinfo {year} {1983})}\BibitemShut
  {NoStop}%
\bibitem [{\citenamefont {Fukuyama}\ \emph {et~al.}(1984)\citenamefont
  {Fukuyama}, \citenamefont {Ebisawa},\ and\ \citenamefont
  {Maekawa}}]{weakly_localized_regime}%
  \BibitemOpen
  \bibfield  {author} {\bibinfo {author} {\bibfnamefont {Hidetoshi}\
  \bibnamefont {Fukuyama}}, \bibinfo {author} {\bibfnamefont {Hiromichi}\
  \bibnamefont {Ebisawa}}, \ and\ \bibinfo {author} {\bibfnamefont {Sadamichi}\
  \bibnamefont {Maekawa}},\ }\bibfield  {title} {\enquote {\bibinfo {title}
  {Bulk superconductivity in weakly localized regime},}\ }\href {\doibase
  10.1143/JPSJ.53.3560} {\bibfield  {journal} {\bibinfo  {journal} {Journal of
  the Physical Society of Japan}\ }\textbf {\bibinfo {volume} {53}},\ \bibinfo
  {pages} {3560--3567} (\bibinfo {year} {1984})}\BibitemShut {NoStop}%
\bibitem [{\citenamefont {Imry}\ and\ \citenamefont
  {Strongin}(1981)}]{destruction_granular}%
  \BibitemOpen
  \bibfield  {author} {\bibinfo {author} {\bibfnamefont {Yoseph}\ \bibnamefont
  {Imry}}\ and\ \bibinfo {author} {\bibfnamefont {Myron}\ \bibnamefont
  {Strongin}},\ }\bibfield  {title} {\enquote {\bibinfo {title} {Destruction of
  superconductivity in granular and highly disordered metals},}\ }\href
  {\doibase 10.1103/PhysRevB.24.6353} {\bibfield  {journal} {\bibinfo
  {journal} {Phys. Rev. B}\ }\textbf {\bibinfo {volume} {24}},\ \bibinfo
  {pages} {6353--6360} (\bibinfo {year} {1981})}\BibitemShut {NoStop}%
\bibitem [{\citenamefont {Strongin}\ \emph {et~al.}(1970)\citenamefont
  {Strongin}, \citenamefont {Thompson}, \citenamefont {Kammerer},\ and\
  \citenamefont {Crow}}]{destruction_films}%
  \BibitemOpen
  \bibfield  {author} {\bibinfo {author} {\bibfnamefont {Myron}\ \bibnamefont
  {Strongin}}, \bibinfo {author} {\bibfnamefont {R.~S.}\ \bibnamefont
  {Thompson}}, \bibinfo {author} {\bibfnamefont {O.~F.}\ \bibnamefont
  {Kammerer}}, \ and\ \bibinfo {author} {\bibfnamefont {J.~E.}\ \bibnamefont
  {Crow}},\ }\bibfield  {title} {\enquote {\bibinfo {title} {Destruction of
  superconductivity in disordered near-monolayer films},}\ }\href {\doibase
  10.1103/PhysRevB.1.1078} {\bibfield  {journal} {\bibinfo  {journal} {Phys.
  Rev. B}\ }\textbf {\bibinfo {volume} {1}},\ \bibinfo {pages} {1078--1091}
  (\bibinfo {year} {1970})}\BibitemShut {NoStop}%
\bibitem [{\citenamefont {Mishonov}\ \emph {et~al.}(2003)\citenamefont
  {Mishonov}, \citenamefont {Penev}, \citenamefont {Indekeu},\ and\
  \citenamefont {Pokrovsky}}]{heat-prb}%
  \BibitemOpen
  \bibfield  {author} {\bibinfo {author} {\bibfnamefont {Todor~M.}\
  \bibnamefont {Mishonov}}, \bibinfo {author} {\bibfnamefont {Evgeni~S.}\
  \bibnamefont {Penev}}, \bibinfo {author} {\bibfnamefont {Joseph~O.}\
  \bibnamefont {Indekeu}}, \ and\ \bibinfo {author} {\bibfnamefont {Valery~L.}\
  \bibnamefont {Pokrovsky}},\ }\bibfield  {title} {\enquote {\bibinfo {title}
  {Specific-heat discontinuity in impure two-band superconductors},}\ }\href
  {\doibase 10.1103/PhysRevB.68.104517} {\bibfield  {journal} {\bibinfo
  {journal} {Phys. Rev. B}\ }\textbf {\bibinfo {volume} {68}},\ \bibinfo
  {pages} {104517} (\bibinfo {year} {2003})}\BibitemShut {NoStop}%
\bibitem [{\citenamefont {Paul}\ \emph {et~al.}(2019)\citenamefont {Paul},
  \citenamefont {Chandra},\ and\ \citenamefont
  {Chattopadhyay}}]{renormalization_elph}%
  \BibitemOpen
  \bibfield  {author} {\bibinfo {author} {\bibfnamefont {Sabyasachi}\
  \bibnamefont {Paul}}, \bibinfo {author} {\bibfnamefont {L~S~Sharath}\
  \bibnamefont {Chandra}}, \ and\ \bibinfo {author} {\bibfnamefont {M~K}\
  \bibnamefont {Chattopadhyay}},\ }\bibfield  {title} {\enquote {\bibinfo
  {title} {Renormalization of electron–phonon coupling in the
  mott–ioffe–regel limit due to point defects in the v$_{1-x}$ti$_x$ alloy
  superconductors},}\ }\href {\doibase 10.1088/1361-648X/ab3515} {\bibfield
  {journal} {\bibinfo  {journal} {Journal of Physics: Condensed Matter}\
  }\textbf {\bibinfo {volume} {31}},\ \bibinfo {pages} {475801} (\bibinfo
  {year} {2019})}\BibitemShut {NoStop}%
\bibitem [{\citenamefont {Jasiewicz}\ \emph {et~al.}(2016)\citenamefont
  {Jasiewicz}, \citenamefont {Wiendlocha}, \citenamefont {Korbe{\'n}},
  \citenamefont {Kaprzyk},\ and\ \citenamefont {Tobola}}]{jasiewicz_2016}%
  \BibitemOpen
  \bibfield  {author} {\bibinfo {author} {\bibfnamefont {K.}~\bibnamefont
  {Jasiewicz}}, \bibinfo {author} {\bibfnamefont {B.}~\bibnamefont
  {Wiendlocha}}, \bibinfo {author} {\bibfnamefont {P.}~\bibnamefont
  {Korbe{\'n}}}, \bibinfo {author} {\bibfnamefont {S.}~\bibnamefont {Kaprzyk}},
  \ and\ \bibinfo {author} {\bibfnamefont {J.}~\bibnamefont {Tobola}},\
  }\bibfield  {title} {\enquote {\bibinfo {title} {{Superconductivity of
  Ta$_{34}$Nb$_{33}$Hf$_8$Zr$_{14}$Ti$_{11}$ high entropy alloy from first
  principles calculations}},}\ }\href {\doibase 10.1002/pssr.201600056}
  {\bibfield  {journal} {\bibinfo  {journal} {Physica Status Solidi (RRL) -
  Rapid Research Letters}\ }\textbf {\bibinfo {volume} {10}},\ \bibinfo {pages}
  {415--419} (\bibinfo {year} {2016})}\BibitemShut {NoStop}%
\bibitem [{\citenamefont {Butler}(1985)}]{butler_1985}%
  \BibitemOpen
  \bibfield  {author} {\bibinfo {author} {\bibfnamefont {W.~H.}\ \bibnamefont
  {Butler}},\ }\bibfield  {title} {\enquote {\bibinfo {title} {Theory of
  electronic transport in random alloys: Korringa-kohn-rostoker
  coherent-potential approximation},}\ }\href {\doibase
  10.1103/PhysRevB.31.3260} {\bibfield  {journal} {\bibinfo  {journal} {Phys.
  Rev. B}\ }\textbf {\bibinfo {volume} {31}},\ \bibinfo {pages} {3260--3277}
  (\bibinfo {year} {1985})}\BibitemShut {NoStop}%
\bibitem [{\citenamefont {{K{\"o}rmann}}\ \emph {et~al.}(2017)\citenamefont
  {{K{\"o}rmann}}, \citenamefont {{Ikeda}}, \citenamefont {{Grabowski}},\ and\
  \citenamefont {{Sluiter}}}]{phonon_broadening}%
  \BibitemOpen
  \bibfield  {author} {\bibinfo {author} {\bibfnamefont {Fritz}\ \bibnamefont
  {{K{\"o}rmann}}}, \bibinfo {author} {\bibfnamefont {Yuji}\ \bibnamefont
  {{Ikeda}}}, \bibinfo {author} {\bibfnamefont {Blazej}\ \bibnamefont
  {{Grabowski}}}, \ and\ \bibinfo {author} {\bibfnamefont {Marcel H.~F.}\
  \bibnamefont {{Sluiter}}},\ }\bibfield  {title} {\enquote {\bibinfo {title}
  {{Phonon broadening in high entropy alloys}},}\ }\href {\doibase
  10.1038/s41524-017-0037-8} {\bibfield  {journal} {\bibinfo  {journal} {npj
  Computational Mathematics}\ }\textbf {\bibinfo {volume} {3}},\ \bibinfo {eid}
  {36} (\bibinfo {year} {2017})}\BibitemShut {NoStop}%
\bibitem [{\citenamefont {Klemens}(1955)}]{Klemens_1955}%
  \BibitemOpen
  \bibfield  {author} {\bibinfo {author} {\bibfnamefont {P~G}\ \bibnamefont
  {Klemens}},\ }\bibfield  {title} {\enquote {\bibinfo {title} {The scattering
  of low-frequency lattice waves by static imperfections},}\ }\href {\doibase
  10.1088/0370-1298/68/12/303} {\bibfield  {journal} {\bibinfo  {journal}
  {Proceedings of the Physical Society. Section A}\ }\textbf {\bibinfo {volume}
  {68}},\ \bibinfo {pages} {1113} (\bibinfo {year} {1955})}\BibitemShut
  {NoStop}%
\bibitem [{\citenamefont {Guo}\ \emph {et~al.}(2017)\citenamefont {Guo},
  \citenamefont {Wang}, \citenamefont {von Rohr}, \citenamefont {Wang},
  \citenamefont {Cai}, \citenamefont {Zhou}, \citenamefont {Yang},
  \citenamefont {Li}, \citenamefont {Jiang}, \citenamefont {Wu}, \citenamefont
  {Cava},\ and\ \citenamefont {Sun}}]{guo-pressure}%
  \BibitemOpen
  \bibfield  {author} {\bibinfo {author} {\bibfnamefont {Jing}\ \bibnamefont
  {Guo}}, \bibinfo {author} {\bibfnamefont {Honghong}\ \bibnamefont {Wang}},
  \bibinfo {author} {\bibfnamefont {Fabian}\ \bibnamefont {von Rohr}}, \bibinfo
  {author} {\bibfnamefont {Zhe}\ \bibnamefont {Wang}}, \bibinfo {author}
  {\bibfnamefont {Shu}\ \bibnamefont {Cai}}, \bibinfo {author} {\bibfnamefont
  {Yazhou}\ \bibnamefont {Zhou}}, \bibinfo {author} {\bibfnamefont
  {Ke}~\bibnamefont {Yang}}, \bibinfo {author} {\bibfnamefont {Aiguo}\
  \bibnamefont {Li}}, \bibinfo {author} {\bibfnamefont {Sheng}\ \bibnamefont
  {Jiang}}, \bibinfo {author} {\bibfnamefont {Qi}~\bibnamefont {Wu}}, \bibinfo
  {author} {\bibfnamefont {Robert~J.}\ \bibnamefont {Cava}}, \ and\ \bibinfo
  {author} {\bibfnamefont {Liling}\ \bibnamefont {Sun}},\ }\bibfield  {title}
  {\enquote {\bibinfo {title} {{Robust zero resistance in a superconducting
  high-entropy alloy at pressures up to 190 GPa}},}\ }\href {\doibase
  10.1073/pnas.1716981114} {\bibfield  {journal} {\bibinfo  {journal}
  {Proceedings of the National Academy of Sciences}\ }\textbf {\bibinfo
  {volume} {114}},\ \bibinfo {pages} {13144--13147} (\bibinfo {year}
  {2017})}\BibitemShut {NoStop}%
\bibitem [{\citenamefont {Jasiewicz}\ \emph {et~al.}(2019)\citenamefont
  {Jasiewicz}, \citenamefont {Wiendlocha}, \citenamefont {G\'ornicka},
  \citenamefont {Gofryk}, \citenamefont {Gazda}, \citenamefont {Klimczuk},\
  and\ \citenamefont {Tobola}}]{jasiewicz2019}%
  \BibitemOpen
  \bibfield  {author} {\bibinfo {author} {\bibfnamefont {K.}~\bibnamefont
  {Jasiewicz}}, \bibinfo {author} {\bibfnamefont {B.}~\bibnamefont
  {Wiendlocha}}, \bibinfo {author} {\bibfnamefont {K.}~\bibnamefont
  {G\'ornicka}}, \bibinfo {author} {\bibfnamefont {K.}~\bibnamefont {Gofryk}},
  \bibinfo {author} {\bibfnamefont {M.}~\bibnamefont {Gazda}}, \bibinfo
  {author} {\bibfnamefont {T.}~\bibnamefont {Klimczuk}}, \ and\ \bibinfo
  {author} {\bibfnamefont {J.}~\bibnamefont {Tobola}},\ }\bibfield  {title}
  {\enquote {\bibinfo {title} {Pressure effects on the electronic structure and
  superconductivity of ${(\mathrm{TaNb})}_{0.67}{(\mathrm{HfZrTi})}_{0.33}$
  high entropy alloy},}\ }\href {\doibase 10.1103/PhysRevB.100.184503}
  {\bibfield  {journal} {\bibinfo  {journal} {Phys. Rev. B}\ }\textbf {\bibinfo
  {volume} {100}},\ \bibinfo {pages} {184503} (\bibinfo {year}
  {2019})}\BibitemShut {NoStop}%
\bibitem [{\citenamefont {Marik}\ \emph {et~al.}(2019)\citenamefont {Marik},
  \citenamefont {Motla}, \citenamefont {Varghese}, \citenamefont {Sajilesh},
  \citenamefont {Singh}, \citenamefont {Breard}, \citenamefont {Boullay},\ and\
  \citenamefont {Singh}}]{marik2019_hex}%
  \BibitemOpen
  \bibfield  {author} {\bibinfo {author} {\bibfnamefont {Sourav}\ \bibnamefont
  {Marik}}, \bibinfo {author} {\bibfnamefont {Kapil}\ \bibnamefont {Motla}},
  \bibinfo {author} {\bibfnamefont {Maneesha}\ \bibnamefont {Varghese}},
  \bibinfo {author} {\bibfnamefont {K.~P.}\ \bibnamefont {Sajilesh}}, \bibinfo
  {author} {\bibfnamefont {Deepak}\ \bibnamefont {Singh}}, \bibinfo {author}
  {\bibfnamefont {Y.}~\bibnamefont {Breard}}, \bibinfo {author} {\bibfnamefont
  {P.}~\bibnamefont {Boullay}}, \ and\ \bibinfo {author} {\bibfnamefont
  {R.~P.}\ \bibnamefont {Singh}},\ }\bibfield  {title} {\enquote {\bibinfo
  {title} {Superconductivity in a new hexagonal high-entropy alloy},}\ }\href
  {\doibase 10.1103/PhysRevMaterials.3.060602} {\bibfield  {journal} {\bibinfo
  {journal} {Phys. Rev. Materials}\ }\textbf {\bibinfo {volume} {3}},\ \bibinfo
  {pages} {060602} (\bibinfo {year} {2019})}\BibitemShut {NoStop}%
\bibitem [{\citenamefont {Mizuguchi}\ \emph {et~al.}(2021)\citenamefont
  {Mizuguchi}, \citenamefont {Kasem},\ and\ \citenamefont
  {Matsuda}}]{mizuguchi2021}%
  \BibitemOpen
  \bibfield  {author} {\bibinfo {author} {\bibfnamefont {Yoshikazu}\
  \bibnamefont {Mizuguchi}}, \bibinfo {author} {\bibfnamefont {Md~Riad}\
  \bibnamefont {Kasem}}, \ and\ \bibinfo {author} {\bibfnamefont {Tatsuma~D}\
  \bibnamefont {Matsuda}},\ }\bibfield  {title} {\enquote {\bibinfo {title}
  {{Superconductivity in CuAl$_2$-type
  Co$_{0.2}$Ni$_{0.1}$Cu$_{0.1}$Rh$_{0.3}$Ir$_{0.3}$Zr$_2$ with a
  high-entropy-alloy transition metal site}},}\ }\href {\doibase
  10.1080/21663831.2020.1860147} {\bibfield  {journal} {\bibinfo  {journal}
  {Materials Research Letters}\ }\textbf {\bibinfo {volume} {9}},\ \bibinfo
  {pages} {141--147} (\bibinfo {year} {2021})}\BibitemShut {NoStop}%
\bibitem [{\citenamefont {Kasem}\ \emph {et~al.}(2021)\citenamefont {Kasem},
  \citenamefont {Yamashita}, \citenamefont {Goto}, \citenamefont {Matsuda},\
  and\ \citenamefont {Mizuguchi}}]{kasem2021}%
  \BibitemOpen
  \bibfield  {author} {\bibinfo {author} {\bibfnamefont {Md~Riad}\ \bibnamefont
  {Kasem}}, \bibinfo {author} {\bibfnamefont {Aichi}\ \bibnamefont
  {Yamashita}}, \bibinfo {author} {\bibfnamefont {Yosuke}\ \bibnamefont
  {Goto}}, \bibinfo {author} {\bibfnamefont {Tatsuma~D}\ \bibnamefont
  {Matsuda}}, \ and\ \bibinfo {author} {\bibfnamefont {Yoshikazu}\ \bibnamefont
  {Mizuguchi}},\ }\bibfield  {title} {\enquote {\bibinfo {title} {{Synthesis of
  high-entropy-alloy-type superconductors (Fe, Co, Ni, Rh, Ir)Zr$_2$ with
  tunable transition temperature}},}\ }\href {\doibase
  10.1007/s10853-021-05921-2} {\bibfield  {journal} {\bibinfo  {journal}
  {Journal of Materials Science}\ }\textbf {\bibinfo {volume} {56}},\ \bibinfo
  {pages} {9499--9505} (\bibinfo {year} {2021})}\BibitemShut {NoStop}%
\bibitem [{\citenamefont {Ioffe}\ and\ \citenamefont
  {Regel}(1960)}]{ioffe1960non}%
  \BibitemOpen
  \bibfield  {author} {\bibinfo {author} {\bibfnamefont {AF}~\bibnamefont
  {Ioffe}}\ and\ \bibinfo {author} {\bibfnamefont {AR}~\bibnamefont {Regel}},\
  }\bibfield  {title} {\enquote {\bibinfo {title} {Non-crystalline, amorphous,
  and liquid electronic semiconductors},}\ }in\ \href@noop {} {\emph {\bibinfo
  {booktitle} {Progress in semiconductors}}}\ (\bibinfo {year} {1960})\ pp.\
  \bibinfo {pages} {237--291}\BibitemShut {NoStop}%
\bibitem [{\citenamefont {‖}\ \emph {et~al.}(2004)\citenamefont {‖},
  \citenamefont {Takenaka},\ and\ \citenamefont
  {Takagi}}]{hussey2004universality}%
  \BibitemOpen
  \bibfield  {author} {\bibinfo {author} {\bibfnamefont {N.~E.~Hussey}\
  \bibnamefont {‖}}, \bibinfo {author} {\bibfnamefont {K.}~\bibnamefont
  {Takenaka}}, \ and\ \bibinfo {author} {\bibfnamefont {H.}~\bibnamefont
  {Takagi}},\ }\bibfield  {title} {\enquote {\bibinfo {title} {{Universality of
  the Mott–Ioffe–Regel limit in metals}},}\ }\href {\doibase
  10.1080/14786430410001716944} {\bibfield  {journal} {\bibinfo  {journal}
  {Philosophical Magazine}\ }\textbf {\bibinfo {volume} {84}},\ \bibinfo
  {pages} {2847--2864} (\bibinfo {year} {2004})}\BibitemShut {NoStop}%
\bibitem [{\citenamefont {Soven}(1967)}]{soven_1967}%
  \BibitemOpen
  \bibfield  {author} {\bibinfo {author} {\bibfnamefont {Paul}\ \bibnamefont
  {Soven}},\ }\bibfield  {title} {\enquote {\bibinfo {title}
  {Coherent-potential model of substitutional disordered alloys},}\ }\href
  {\doibase 10.1103/PhysRev.156.809} {\bibfield  {journal} {\bibinfo  {journal}
  {Phys. Rev.}\ }\textbf {\bibinfo {volume} {156}},\ \bibinfo {pages}
  {809--813} (\bibinfo {year} {1967})}\BibitemShut {NoStop}%
\bibitem [{\citenamefont {Kaprzyk}\ and\ \citenamefont
  {Bansil}(1990)}]{kaprzyk_1990}%
  \BibitemOpen
  \bibfield  {author} {\bibinfo {author} {\bibfnamefont {S.}~\bibnamefont
  {Kaprzyk}}\ and\ \bibinfo {author} {\bibfnamefont {A.}~\bibnamefont
  {Bansil}},\ }\bibfield  {title} {\enquote {\bibinfo {title} {Green's function
  and a generalized lloyd formula for the density of states in disordered
  muffin-tin alloys},}\ }\href {\doibase 10.1103/PhysRevB.42.7358} {\bibfield
  {journal} {\bibinfo  {journal} {Phys. Rev. B}\ }\textbf {\bibinfo {volume}
  {42}},\ \bibinfo {pages} {7358--7362} (\bibinfo {year} {1990})}\BibitemShut
  {NoStop}%
\bibitem [{\citenamefont {Bansil}\ \emph {et~al.}(1999)\citenamefont {Bansil},
  \citenamefont {Kaprzyk}, \citenamefont {Mijnarends},\ and\ \citenamefont
  {Tobo\l{}a}}]{bansil_1999}%
  \BibitemOpen
  \bibfield  {author} {\bibinfo {author} {\bibfnamefont {A.}~\bibnamefont
  {Bansil}}, \bibinfo {author} {\bibfnamefont {S.}~\bibnamefont {Kaprzyk}},
  \bibinfo {author} {\bibfnamefont {P.~E.}\ \bibnamefont {Mijnarends}}, \ and\
  \bibinfo {author} {\bibfnamefont {J.}~\bibnamefont {Tobo\l{}a}},\ }\bibfield
  {title} {\enquote {\bibinfo {title} {{Electronic structure and magnetism of
  ${\mathrm{Fe}}_{3\ensuremath{-}x}{\mathrm{V}}_{x}X$ $(X=\mathrm{Si},$ Ga, and
  Al) alloys by the KKR-CPA method}},}\ }\href {\doibase
  10.1103/PhysRevB.60.13396} {\bibfield  {journal} {\bibinfo  {journal} {Phys.
  Rev. B}\ }\textbf {\bibinfo {volume} {60}},\ \bibinfo {pages} {13396--13412}
  (\bibinfo {year} {1999})}\BibitemShut {NoStop}%
\bibitem [{\citenamefont {Stopa}\ \emph {et~al.}(2004)\citenamefont {Stopa},
  \citenamefont {Kaprzyk},\ and\ \citenamefont {Tobola}}]{stopa_2004}%
  \BibitemOpen
  \bibfield  {author} {\bibinfo {author} {\bibfnamefont {T}~\bibnamefont
  {Stopa}}, \bibinfo {author} {\bibfnamefont {S}~\bibnamefont {Kaprzyk}}, \
  and\ \bibinfo {author} {\bibfnamefont {J}~\bibnamefont {Tobola}},\ }\bibfield
   {title} {\enquote {\bibinfo {title} {Linear aspects of the
  korringa{\textendash}kohn{\textendash}rostoker formalism},}\ }\href {\doibase
  10.1088/0953-8984/16/28/012} {\bibfield  {journal} {\bibinfo  {journal}
  {Journal of Physics: Condensed Matter}\ }\textbf {\bibinfo {volume} {16}},\
  \bibinfo {pages} {4921--4933} (\bibinfo {year} {2004})}\BibitemShut {NoStop}%
\bibitem [{\citenamefont {Perdew}\ and\ \citenamefont
  {Wang}(1992)}]{perdew_1992}%
  \BibitemOpen
  \bibfield  {author} {\bibinfo {author} {\bibfnamefont {John~P.}\ \bibnamefont
  {Perdew}}\ and\ \bibinfo {author} {\bibfnamefont {Yue}\ \bibnamefont
  {Wang}},\ }\bibfield  {title} {\enquote {\bibinfo {title} {Accurate and
  simple analytic representation of the electron-gas correlation energy},}\
  }\href {\doibase 10.1103/PhysRevB.45.13244} {\bibfield  {journal} {\bibinfo
  {journal} {Phys. Rev. B}\ }\textbf {\bibinfo {volume} {45}},\ \bibinfo
  {pages} {13244--13249} (\bibinfo {year} {1992})}\BibitemShut {NoStop}%
\bibitem [{\citenamefont {Gaspari}\ and\ \citenamefont
  {Gyorffy}(1972)}]{gaspari_1972}%
  \BibitemOpen
  \bibfield  {author} {\bibinfo {author} {\bibfnamefont {G.~D.}\ \bibnamefont
  {Gaspari}}\ and\ \bibinfo {author} {\bibfnamefont {B.~L.}\ \bibnamefont
  {Gyorffy}},\ }\bibfield  {title} {\enquote {\bibinfo {title}
  {{Electron-Phonon Interactions, $d$ Resonances, and Superconductivity in
  Transition Metals}},}\ }\href {\doibase 10.1103/PhysRevLett.28.801}
  {\bibfield  {journal} {\bibinfo  {journal} {Phys. Rev. Lett.}\ }\textbf
  {\bibinfo {volume} {28}},\ \bibinfo {pages} {801--805} (\bibinfo {year}
  {1972})}\BibitemShut {NoStop}%
\bibitem [{\citenamefont {Gomersall}\ and\ \citenamefont
  {Gyorffy}(1974)}]{gomersall_1974}%
  \BibitemOpen
  \bibfield  {author} {\bibinfo {author} {\bibfnamefont {I~R}\ \bibnamefont
  {Gomersall}}\ and\ \bibinfo {author} {\bibfnamefont {B~L}\ \bibnamefont
  {Gyorffy}},\ }\bibfield  {title} {\enquote {\bibinfo {title} {A simple theory
  of the electron-phonon mass enhancement in transition metal compounds},}\
  }\href {\doibase 10.1088/0305-4608/4/8/015} {\bibfield  {journal} {\bibinfo
  {journal} {Journal of Physics F: Metal Physics}\ }\textbf {\bibinfo {volume}
  {4}},\ \bibinfo {pages} {1204--1221} (\bibinfo {year} {1974})}\BibitemShut
  {NoStop}%
\bibitem [{\citenamefont {Klein}\ \emph {et~al.}(1979)\citenamefont {Klein},
  \citenamefont {Boyer},\ and\ \citenamefont {Papaconstantopoulos}}]{papa_a15}%
  \BibitemOpen
  \bibfield  {author} {\bibinfo {author} {\bibfnamefont {B.~M.}\ \bibnamefont
  {Klein}}, \bibinfo {author} {\bibfnamefont {L.~L.}\ \bibnamefont {Boyer}}, \
  and\ \bibinfo {author} {\bibfnamefont {D.~A.}\ \bibnamefont
  {Papaconstantopoulos}},\ }\bibfield  {title} {\enquote {\bibinfo {title}
  {{Superconducting Properties of $A15$ Compounds Derived from Band-Structure
  Results}},}\ }\href {\doibase 10.1103/PhysRevLett.42.530} {\bibfield
  {journal} {\bibinfo  {journal} {Phys. Rev. Lett.}\ }\textbf {\bibinfo
  {volume} {42}},\ \bibinfo {pages} {530--533} (\bibinfo {year}
  {1979})}\BibitemShut {NoStop}%
\bibitem [{\citenamefont {Mazin}\ \emph {et~al.}(1990)\citenamefont {Mazin},
  \citenamefont {Rashkeev},\ and\ \citenamefont {Savrasov}}]{mazin_1990}%
  \BibitemOpen
  \bibfield  {author} {\bibinfo {author} {\bibfnamefont {I.~I.}\ \bibnamefont
  {Mazin}}, \bibinfo {author} {\bibfnamefont {S.~N.}\ \bibnamefont {Rashkeev}},
  \ and\ \bibinfo {author} {\bibfnamefont {S.~Y.}\ \bibnamefont {Savrasov}},\
  }\bibfield  {title} {\enquote {\bibinfo {title} {{Nonspherical
  rigid-muffin-tin calculations of electron-phonon coupling in
  high-${\mathit{T}}_{\mathit{c}}$ perovskites}},}\ }\href {\doibase
  10.1103/PhysRevB.42.366} {\bibfield  {journal} {\bibinfo  {journal} {Phys.
  Rev. B}\ }\textbf {\bibinfo {volume} {42}},\ \bibinfo {pages} {366--370}
  (\bibinfo {year} {1990})}\BibitemShut {NoStop}%
\bibitem [{\citenamefont {Wiendlocha}\ \emph {et~al.}(2006)\citenamefont
  {Wiendlocha}, \citenamefont {Tobola},\ and\ \citenamefont
  {Kaprzyk}}]{wiendlocha_2006}%
  \BibitemOpen
  \bibfield  {author} {\bibinfo {author} {\bibfnamefont {B.}~\bibnamefont
  {Wiendlocha}}, \bibinfo {author} {\bibfnamefont {J.}~\bibnamefont {Tobola}},
  \ and\ \bibinfo {author} {\bibfnamefont {S.}~\bibnamefont {Kaprzyk}},\
  }\bibfield  {title} {\enquote {\bibinfo {title} {{Search for
  ${\mathrm{Sc}}_{3}X\mathrm{B}$
  $(X=\mathrm{In},\mathrm{Tl},\mathrm{Ga},\mathrm{Al})$ perovskites
  superconductors and proximity of weak ferromagnetism}},}\ }\href {\doibase
  10.1103/PhysRevB.73.134522} {\bibfield  {journal} {\bibinfo  {journal} {Phys.
  Rev. B}\ }\textbf {\bibinfo {volume} {73}},\ \bibinfo {pages} {134522}
  (\bibinfo {year} {2006})}\BibitemShut {NoStop}%
\bibitem [{\citenamefont {Wiendlocha}\ \emph {et~al.}(2008)\citenamefont
  {Wiendlocha}, \citenamefont {Tobola}, \citenamefont {Sternik}, \citenamefont
  {Kaprzyk}, \citenamefont {Parlinski},\ and\ \citenamefont
  {Ole\ifmmode~\acute{s}\else \'{s}\fi{}}}]{wiendlocha2008}%
  \BibitemOpen
  \bibfield  {author} {\bibinfo {author} {\bibfnamefont {B.}~\bibnamefont
  {Wiendlocha}}, \bibinfo {author} {\bibfnamefont {J.}~\bibnamefont {Tobola}},
  \bibinfo {author} {\bibfnamefont {M.}~\bibnamefont {Sternik}}, \bibinfo
  {author} {\bibfnamefont {S.}~\bibnamefont {Kaprzyk}}, \bibinfo {author}
  {\bibfnamefont {K.}~\bibnamefont {Parlinski}}, \ and\ \bibinfo {author}
  {\bibfnamefont {A.~M.}\ \bibnamefont {Ole\ifmmode~\acute{s}\else
  \'{s}\fi{}}},\ }\bibfield  {title} {\enquote {\bibinfo {title}
  {{Superconductivity of ${\text{Mo}}_{3}{\text{Sb}}_{7}$ from first
  principles}},}\ }\href {\doibase 10.1103/PhysRevB.78.060507} {\bibfield
  {journal} {\bibinfo  {journal} {Phys. Rev. B}\ }\textbf {\bibinfo {volume}
  {78}},\ \bibinfo {pages} {060507(R)} (\bibinfo {year} {2008})}\BibitemShut
  {NoStop}%
\bibitem [{\citenamefont {Wiendlocha}\ and\ \citenamefont
  {Sternik}(2014)}]{wiendlocha2014}%
  \BibitemOpen
  \bibfield  {author} {\bibinfo {author} {\bibfnamefont {Bartlomiej}\
  \bibnamefont {Wiendlocha}}\ and\ \bibinfo {author} {\bibfnamefont
  {Malgorzata}\ \bibnamefont {Sternik}},\ }\bibfield  {title} {\enquote
  {\bibinfo {title} {{Effect of the tetragonal distortion on the electronic
  structure, phonons and superconductivity in the
  ${\text{Mo}}_{3}{\text{Sb}}_{7}$ superconductor}},}\ }\href {\doibase
  https://doi.org/10.1016/j.intermet.2014.05.002} {\bibfield  {journal}
  {\bibinfo  {journal} {Intermetallics}\ }\textbf {\bibinfo {volume} {53}},\
  \bibinfo {pages} {150 -- 156} (\bibinfo {year} {2014})}\BibitemShut {NoStop}%
\bibitem [{\citenamefont {Rajput}\ \emph {et~al.}(1996)\citenamefont {Rajput},
  \citenamefont {Prasad}, \citenamefont {Singru}, \citenamefont {Kaprzyk},\
  and\ \citenamefont {Bansil}}]{kaprzyk_1996}%
  \BibitemOpen
  \bibfield  {author} {\bibinfo {author} {\bibfnamefont {S~S}\ \bibnamefont
  {Rajput}}, \bibinfo {author} {\bibfnamefont {R}~\bibnamefont {Prasad}},
  \bibinfo {author} {\bibfnamefont {R~M}\ \bibnamefont {Singru}}, \bibinfo
  {author} {\bibfnamefont {S}~\bibnamefont {Kaprzyk}}, \ and\ \bibinfo {author}
  {\bibfnamefont {A}~\bibnamefont {Bansil}},\ }\bibfield  {title} {\enquote
  {\bibinfo {title} {{Electronic structure of disordered Nb - Mo alloys studied
  using the charge-self-consistent Korringa - Kohn - Rostoker coherent
  potential approximation}},}\ }\href {\doibase 10.1088/0953-8984/8/17/006}
  {\bibfield  {journal} {\bibinfo  {journal} {Journal of Physics: Condensed
  Matter}\ }\textbf {\bibinfo {volume} {8}},\ \bibinfo {pages} {2929--2944}
  (\bibinfo {year} {1996})}\BibitemShut {NoStop}%
\bibitem [{\citenamefont {Berk}\ and\ \citenamefont {Schrieffer}(1966)}]{berk}%
  \BibitemOpen
  \bibfield  {author} {\bibinfo {author} {\bibfnamefont {N.~F.}\ \bibnamefont
  {Berk}}\ and\ \bibinfo {author} {\bibfnamefont {J.~R.}\ \bibnamefont
  {Schrieffer}},\ }\bibfield  {title} {\enquote {\bibinfo {title} {Effect of
  ferromagnetic spin correlations on superconductivity},}\ }\href {\doibase
  10.1103/PhysRevLett.17.433} {\bibfield  {journal} {\bibinfo  {journal} {Phys.
  Rev. Lett.}\ }\textbf {\bibinfo {volume} {17}},\ \bibinfo {pages} {433--435}
  (\bibinfo {year} {1966})}\BibitemShut {NoStop}%
\bibitem [{\citenamefont {Webb}\ \emph {et~al.}(1978)\citenamefont {Webb},
  \citenamefont {Ketterson}, \citenamefont {Halperin}, \citenamefont
  {Vuillemin},\ and\ \citenamefont {Sandesara}}]{pd}%
  \BibitemOpen
  \bibfield  {author} {\bibinfo {author} {\bibfnamefont {RA}~\bibnamefont
  {Webb}}, \bibinfo {author} {\bibfnamefont {JB}~\bibnamefont {Ketterson}},
  \bibinfo {author} {\bibfnamefont {WP}~\bibnamefont {Halperin}}, \bibinfo
  {author} {\bibfnamefont {JJ}~\bibnamefont {Vuillemin}}, \ and\ \bibinfo
  {author} {\bibfnamefont {NB}~\bibnamefont {Sandesara}},\ }\bibfield  {title}
  {\enquote {\bibinfo {title} {{Very low temperature search for
  superconductivity in Pd, Pt, and Rh}},}\ }\href {\doibase 10.1007/BF00056650}
  {\bibfield  {journal} {\bibinfo  {journal} {J. Low Temp. Phys.}\ }\textbf
  {\bibinfo {volume} {32}},\ \bibinfo {pages} {659} (\bibinfo {year}
  {1978})}\BibitemShut {NoStop}%
\bibitem [{\citenamefont {Buchal}\ \emph {et~al.}(1983)\citenamefont {Buchal},
  \citenamefont {Pobell}, \citenamefont {Mueller}, \citenamefont {Kubota},\
  and\ \citenamefont {Owers-Bradley}}]{rhodium-superconductivity}%
  \BibitemOpen
  \bibfield  {author} {\bibinfo {author} {\bibfnamefont {Ch.}\ \bibnamefont
  {Buchal}}, \bibinfo {author} {\bibfnamefont {F.}~\bibnamefont {Pobell}},
  \bibinfo {author} {\bibfnamefont {R.~M.}\ \bibnamefont {Mueller}}, \bibinfo
  {author} {\bibfnamefont {M.}~\bibnamefont {Kubota}}, \ and\ \bibinfo {author}
  {\bibfnamefont {J.~R.}\ \bibnamefont {Owers-Bradley}},\ }\bibfield  {title}
  {\enquote {\bibinfo {title} {{Superconductivity of Rhodium at Ultralow
  Temperatures}},}\ }\href {\doibase 10.1103/PhysRevLett.50.64} {\bibfield
  {journal} {\bibinfo  {journal} {Phys. Rev. Lett.}\ }\textbf {\bibinfo
  {volume} {50}},\ \bibinfo {pages} {64--67} (\bibinfo {year}
  {1983})}\BibitemShut {NoStop}%
\bibitem [{\citenamefont {Allen}(1987)}]{allen}%
  \BibitemOpen
  \bibfield  {author} {\bibinfo {author} {\bibfnamefont {Philip~B.}\
  \bibnamefont {Allen}},\ }\bibfield  {title} {\enquote {\bibinfo {title}
  {Empirical electron-phonon $\ensuremath{\lambda}$ values from resistivity of
  cubic metallic elements},}\ }\href {\doibase 10.1103/PhysRevB.36.2920}
  {\bibfield  {journal} {\bibinfo  {journal} {Phys. Rev. B}\ }\textbf {\bibinfo
  {volume} {36}},\ \bibinfo {pages} {2920--2923} (\bibinfo {year}
  {1987})}\BibitemShut {NoStop}%
\bibitem [{\citenamefont {Robarts}\ \emph {et~al.}(2020)\citenamefont
  {Robarts}, \citenamefont {Millichamp}, \citenamefont {Lagos}, \citenamefont
  {Laverock}, \citenamefont {Billington}, \citenamefont {Duffy}, \citenamefont
  {O'Neill}, \citenamefont {Giblin}, \citenamefont {Taylor}, \citenamefont
  {Kontrym-Sznajd}, \citenamefont {Samsel-Czeka\l{}a}, \citenamefont {Bei},
  \citenamefont {Mu}, \citenamefont {Samolyuk}, \citenamefont {Stocks},\ and\
  \citenamefont {Dugdale}}]{extreme_smearing_prl}%
  \BibitemOpen
  \bibfield  {author} {\bibinfo {author} {\bibfnamefont {Hannah~C.}\
  \bibnamefont {Robarts}}, \bibinfo {author} {\bibfnamefont {Thomas~E.}\
  \bibnamefont {Millichamp}}, \bibinfo {author} {\bibfnamefont {Daniel~A.}\
  \bibnamefont {Lagos}}, \bibinfo {author} {\bibfnamefont {Jude}\ \bibnamefont
  {Laverock}}, \bibinfo {author} {\bibfnamefont {David}\ \bibnamefont
  {Billington}}, \bibinfo {author} {\bibfnamefont {Jonathan~A.}\ \bibnamefont
  {Duffy}}, \bibinfo {author} {\bibfnamefont {Daniel}\ \bibnamefont {O'Neill}},
  \bibinfo {author} {\bibfnamefont {Sean~R.}\ \bibnamefont {Giblin}}, \bibinfo
  {author} {\bibfnamefont {Jonathan~W.}\ \bibnamefont {Taylor}}, \bibinfo
  {author} {\bibfnamefont {Grazyna}\ \bibnamefont {Kontrym-Sznajd}}, \bibinfo
  {author} {\bibfnamefont {Ma\l{}gorzata}\ \bibnamefont {Samsel-Czeka\l{}a}},
  \bibinfo {author} {\bibfnamefont {Hongbin}\ \bibnamefont {Bei}}, \bibinfo
  {author} {\bibfnamefont {Sai}\ \bibnamefont {Mu}}, \bibinfo {author}
  {\bibfnamefont {German~D.}\ \bibnamefont {Samolyuk}}, \bibinfo {author}
  {\bibfnamefont {G.~Malcolm}\ \bibnamefont {Stocks}}, \ and\ \bibinfo {author}
  {\bibfnamefont {Stephen~B.}\ \bibnamefont {Dugdale}},\ }\bibfield  {title}
  {\enquote {\bibinfo {title} {Extreme fermi surface smearing in a maximally
  disordered concentrated solid solution},}\ }\href {\doibase
  10.1103/PhysRevLett.124.046402} {\bibfield  {journal} {\bibinfo  {journal}
  {Phys. Rev. Lett.}\ }\textbf {\bibinfo {volume} {124}},\ \bibinfo {pages}
  {046402} (\bibinfo {year} {2020})}\BibitemShut {NoStop}%
\end{thebibliography}%


\end{document}